\documentclass[useAMS,usegraphicx]{mn2e}
\usepackage{rotate}
\usepackage{rotating}
\usepackage{times}
\usepackage{verbatim}
\newif\ifAMStwofonts
\AMStwofontstrue

\def\gs{\mathrel{\hbox{\rlap{\hbox{\lower4pt\hbox{$\sim$}}}\hbox{$>$}}}}
\def\ls{\mathrel{\hbox{\rlap{\hbox{\lower4pt\hbox{$\sim$}}}\hbox{$<$}}}}

\def\suzaku{{\it Suzaku}}

\def\xmm{{\it XMM-Newton}}

\def\nustar{{\it NuSTAR}}

\def\et{{et al.\ }}
\def\mcg{{MCG--6-30-15}}
\def\mrk335{{Mrk~335}}
\def\rg{{\thinspace r_{\rm g}}}
\def\risco{{\thinspace r_{\rm ISCO}}}
\def\fvar{{F_{\rm var}}}
\def\chidof{{\chi^2_\nu/{\rm dof}}}
\def\redchi{{\chi^2_\nu}}
\def\delchi{{\Delta\chi^2}}
\def\feka{{Fe~K$\alpha$}}

\def\fekb{{Fe~K$\beta$}}

\def\nh{{N_{\rm H}}}

\def\deg{^{\circ}}

\def\cm{{\rm\thinspace cm}}
\def\erg{{\rm\thinspace erg}}
\def\eV{{\rm\thinspace eV}}

\def\ga{{\rm\thinspace gauss}}

\def\keV{{\rm\thinspace keV}}
\def\km{{\rm\thinspace km}}

\def\Mpc{{\rm\thinspace Mpc}}
\def\Msun{\hbox{$\rm\thinspace M_{\odot}$}}

\def\s{{\rm\thinspace s}}
\def\ks{{\rm\thinspace ks}}
\def\ps{{\rm\thinspace s^{-1}}}

\def\cts{{\rm\thinspace count}}

\def\cps{\hbox{$\cts\s^{-1}\,$}}
\def\cmps{\hbox{$\cm\s^{-1}\,$}}

\def\ergpscmps{\hbox{$\erg\cm^{-2}\s^{-1}\,$}}
\def\ergcmps{\hbox{$\erg\cm\s^{-1}\,$}}

\def\kmps{\hbox{$\km\ps\,$}}

\def\pscm{\hbox{$\cm^{-2}\,$}}

\title[ \suzaku\ observations of \mrk335]
      {
 \suzaku\ observations of \mrk335: Confronting partial covering and relativistic reflection      }

\author[L. C. Gallo et al.]
       {L. C. Gallo,$^1$ 
       D. R. Wilkins,$^1$
       K. Bonson,$^1$
       C-Y. Chiang,$^1$
       D. Grupe,$^2$
       M. L. Parker,$^3$
       	\newauthor	
       A. Zoghbi,$^4$
       A. C. Fabian,$^3$
	S. Komossa$^5$
	and A. L. Longinotti$^{6,7}$
        \\ 
$^{1}$ Department of Astronomy and Physics, Saint Mary's University, 923 Robie Street, Halifax, NS, B3H 3C3, Canada \\
$^{2}$ Space Science Center, Morehead State University, 235 Martindale Drive, Morehead, KY 40351, USA \\
$^{3}$ Institute of Astronomy, University of Cambridge, Madingley Road, Cambridge CB3 0HA\\
$^{4}$ Department of Astronomy, University of Maryland, College Park, MD 20742, USA \\
$^{5}$  Max-Planck-Institut f\"ur Radioastronomie, Auf dem H\"ugel 69, 53121 Bonn, Germany \\
$^{6}$ XMM-Newton Science Operations Centre, ESA, Villafranca del Castillo, Apartado 78, E-28691 Villanueva de la Ca\~nada, Spain \\
$^{7}$ Departamento de Astronomia Extragalactica y Cosmologia, Instituto de Astronomia, Universidad Nacional Autonoma de Mexico (UNAM), \\
~~ Apartado Postal 70-264, 04510 Mexico D.F., Mexico \\
}
\date{Accepted. Received. }
\pagerange{\pageref{firstpage}--\pageref{lastpage}}
\pubyear{2010}
\begin{document}
\maketitle
\label{firstpage}

\begin{abstract}
We report on the deepest X-ray observation of the narrow-line Seyfert~1 galaxy  \mrk335\  in the low-flux state obtained with \suzaku.  The data are compared to a 2006 high-flux \suzaku\ observation when the source was $\sim 10\times$ brighter.   Describing the two flux levels self-consistently  with partial covering models would require extreme circumstances, as the source would be subject to negligible absorption during the bright state and ninety-five per cent covering with near Compton-thick material when dim.  Blurred reflection from an accretion disc around a nearly maximum spinning black hole ($a>0.91$,  with preference for a spin parameter as high as $\sim 0.995$) appears more likely and is consistent with the long-term and rapid variability.  Measurements of the emissivity profile and spectral modelling indicate the high-flux \suzaku\ observation of \mrk335\ is consistent with continuum-dominated, jet-like emission (i.e. beamed away from the disc).  It can be argued that the ejecta must be confined to within $\sim25\rg$ if it does not escape the system. During the low-flux state the corona becomes compact and only extends to about $5\rg$ from the black hole, and the spectrum becomes reflection-dominated.  The low-frequency lags measured at both epochs are comparable indicating that the accretion mechanism is not changing between the two flux levels.  Various techniques to study the spectral variability (e.g. principal component analysis, fractional variability, difference spectra, and hardness ratio analysis) indicate that the low-state variability is dominated by changes in the power law flux and photon index,  but that changes in the ionisation state of the reflector are also required.   Most notably, the ionisation parameter becomes inversely correlated with the reflected flux after a long-duration flare-like event during the observation.

\end{abstract}



\section{Introduction}
\label{sect:intro}

Partial-covering absorption (e.g. Tanaka \et 2004) and blurred reflection models (e.g. Ross \& Fabian 2005) are consistently debated in attempts to explain the X-ray behaviour of active galactic nuclei (AGN).  X-ray weak and low-flux observations provide the potential to study the underlying physical processes in the absence of the power law continuum that dominates the spectrum at high-flux levels (e.g Gallo 2006).  For example, observing AGN when they are dimmed by extreme absorption can potentially constrain the ionisation, column density and location of the absorber with high precision, providing insight on the material being accreted or expelled by the black hole system (e.g. Turner \& Miller 2009).  On the other hand, observing the reflection dominated spectrum in the low-flux state can reveal the nature of the inner accretion disc (e.g. Reynolds \et 2012), the physics of matter under extreme gravity (e.g. Fabian \et 2009), the geometry of the X-ray region (e.g. Wilkins \et 2014), and even the black hole spin (e.g. Brenneman \& Reynolds 2006).  Consequently, campaigns to catch AGN at low-flux levels have become increasingly important (e.g. Gallo \et 2011a, 2011b; Fabian \et 2012;  Grupe \et 2012, hereafter G12).   The challenge is then to deal with the reduced signal-to-noise that accompanies the low-flux levels.

The narrow-line Seyfert 1 galaxy (NLS1), \mrk335 ($z=0.025785$), has invited much interest in the last decade.  Historically it has been known as one of the brightest X-ray sources in the sky even detected with {\it UHURU} (Tananbaum \et 1978), but  its departure into an x-ray weak and consistent low or moderate flux state (e.g Grupe \et 2008) marks what may be considered the new normal for this active galaxy.   Its lowest flux level in recent years was about $1/30$ of the previously lowest observed flux (Grupe \et 2007), but it is regularly only $\sim1/10$ the expected X-ray flux given its UV brightness  (G12).  Despite the extremely diminished X-ray flux, \mrk335\ remains relatively bright and therefore the source provides a unique opportunity to study the behaviour of AGN at low flux levels with respectable signal-to-noise.  

Data of \mrk335\ when the source was bright have been described comparably well with partial covering (e.g. Tanaka \et 2005; Longinotti \et 2007; O'Neill \et 2007) and blurred reflection (e.g. Ballantyne \et 2001; Crummy \et 2006; Larsson \et 2008).  Timing studies, in particular the lag analysis of the iron K$\alpha$ emission line  in the \xmm\ high state, strongly support the relativistic reflection interpretation (Kara \et 2013).    The source consistently displays a soft excess below $\sim1\keV$ that has been attributed to blurred reflection (e.g. Crummy \et 2006), Comptonisation (e.g. Patrick \et 2011), and absorption (e.g. Middleton \et 2007).  A snap-shot \xmm\ observation of \mrk335\ in the low state could be fitted with partial covering or blurred reflection (Grupe \et 2008), but a broken power law continuum was required when fitting with the partial covering model (Grupe \et 2008).  A longer \xmm\ observation in 2009 caught \mrk335\ at an intermediate flux level as the AGN was transitioning from a low- to high-flux level (G12; Gallo \et 2013, hereafter G13).  G13 demonstrated the spectral and timing results were completely consistent with the blurred reflection interpretation for \mrk335.  Of additional interest with the intermediate flux state observation was that  it was the first time, \mrk335\ exhibited the effects of a warm absorber that was located within the broad line region and likely coming out of the accretion disc (Longinotti \et 2013, hereafter L13).

Although the source flux drops regularly \mrk335\ is highly variable and normally recovers quickly.  The low-flux state of the AGN has remained evasive until now.  These new \suzaku\ observations from 2013 provide the deepest pointing and longest look at  \mrk335\ in the low-flux state.  Observed for $\sim300\ks$ over an approximately $7$~day period these data provide about $8\times$ as many source counts as the 2007 \xmm\ low-flux observation and cover an energy bandpass between $\sim 0.5-40\keV$.  The data provide the opportunity to directly compare the low-flux state with the 2006 \suzaku\ bright state ($\sim150\ks$ exposure) and examine the low-flux behaviour of the AGN in detail.

The observations and data reduction are described in the next section.  In Section~3 the differences between the high-flux and low-flux states are examined as we attempt to describe the spectral changes with both blurred reflection and partial covering.  Section~4 is a detailed examination of the spectral variability in the low-flux state.  Discussion and conclusions following in Section~5 and 6, respectively.

\section{Observations, data reduction, and MCMC error analysis}
\label{sect:data}
\mrk335\ has been observed with \suzaku\ (Mitsuda \et 2007) on two occasions in 2006 and 2013. 
Both observations will be examined here with added emphasis on the 2013 Target of Opportunity observation that was triggered to catch \mrk335\ at a low-flux level.  The 2006 high state was the subject of a detailed analysis by Larsson \et (2008).  The duration of the 2006 and 2013 observations were $\sim3.7$ and $\sim7.7$ days, respectively.  With the approximate $50$ per cent on-target time due to the low-earth orbit of \suzaku, exposures are about half the duration time.   A summary of the observations is provided in Table~\ref{tab:obslog}.  The new low-flux observation provides a considerable improvement ($\approx 8\times$ more source counts) over the 2007 \xmm\ snap-shot that collected only about $13000$ source counts in the $0.7-10\keV$ band.

            \begin{table*}
            \begin{center}
            \caption{\suzaku\ observation log for \mrk335.
            The start date of the observation is given in column (1) and its duration in column (2).   The telescope and instrument used is shown in column (3) and (4), respectively. 
            Column (5) is the ID corresponding to the observation and column (6) is the accumulated good exposure time over the duration of the observation.  Column (7) lists the approximate background subtracted source counts in the $0.7-10\keV$ and $15-40\keV$ band for the XIS and PIN, respectively.  Note the 2006 XIS observation is the summation of three FI chips whereas the 2013 observation includes data from only two chips.  The 2013 observation is continuous, but was divided in to two observational data sets.  
            }
            \begin{tabular}{ccccccc}                
            \hline
            (1) & (2) & (3) & (4) & (5) & (6) & (7) \\
            Start Date   &  Duration &Telescope&   Instrument  & Observation ID    &  Exposure & Source Counts \\
              (year.mm.dd)   &  (Days) &  &    &                    &       (s)    & \\
			\hline
            \hline
            2006.06.21 & 3.7 & \suzaku\  & BI & 70103010  &  132800 &  244055 \\
                & & & FI &             &              132800 & 578116 \\
                 & & & PIN &           &            131700 & 4223  \\
            2013.06.11 & 7.7 & \suzaku\  & BI & 708016010(20)  &  251400 & 42840 \\
                & &  & FI &              &              251450 & 70539 \\
                 & & & PIN &          &            254200 & 7697 \\
            \hline
            \label{tab:obslog}
            \end{tabular}
            \end{center}
            \end{table*}

  During the 2013 \suzaku\ observation the two front-illuminated (FI) CCDs
        (XIS0 and XIS3), the back-illuminated (BI) CCD (XIS1), and the HXD-PIN
        all functioned
        normally and collected data.  The target was observed in the
        XIS-nominal position.  In 2006, the AGN was observed in HXD-nominal position and the third FI CCD (XIS2) was still operational.

        Cleaned event files from version 2 processed data were used in the
        analysis and data
        products were extracted using {\sc xselect}.
        For each XIS chip, source counts were extracted from a $4.3^{\prime}$
        circular region centred on
        the target.  Background counts were taken from surrounding regions on
        the chip excluding the calibration source.  Response files
        (rmf and arf) were
        generated using {\sc xisrmfgen} and {\sc xissimarfgen}.
        After examining for consistency, the data from the XIS-FI were combined
        to create a single spectrum.  Data below $0.7\keV$ and between $1.5-2.5\keV$ are not used for spectral fitting given the 
        existing calibration uncertainties in those bands.  The $7.4-7.8\keV$ band is ignored in the 2013 spectral data due to the strong background emission from Ni~K$\alpha$ compared to the relatively low source flux.  
        
        The PIN spectrum was extracted from the HXD data following standard procedures. We extracted the non-X-ray background (NXB) from the event file obtained directly from the Suzaku Data Center using the tuned model. The data were also corrected for detector deadtime. The cosmic X-ray background (CXB) was modelled using the provided flat response files. We combined the NXB and CXB and created a total background spectrum and found that the source is not detected in the PIN energy band at both epochs. The earth-occulted background, which is ideally consistent with the NXB, was also generated for a further check. We found that the $15-40\keV$ count rate of the earth-occulted background appears lower ($0.194\pm0.002$) than that of the NXB ($0.217\pm0.004$). When the NXB and the earth-occulted background are discrepant, the latter should be used. We then combined the earth-occulted background and the CXB to form a new background. The source is effectively detected between $15-40\keV$ when adopting the new background file and the resulting PIN exposures were $131 \ks$ and $254\ks$ in 2006 and 2013, respective. 

        All parameters are reported in the rest frame of the source unless specified
        otherwise.  The quoted errors on the model parameters correspond to a 90\% confidence level.
        Errors on the model parameters were computed by Markov chain Monte Carlo (MCMC) calculations (see also Reynolds et al 2012, Steiner \& McClintock 2012). The MCMC algorithm computes the probability distribution of the model parameters given the data (i.e. the posterior probability), from which both the best-fitting values and the errors can be determined. The algorithm performs a `guided walk' through the parameter space such that the probability distributions of each of the parameters among the chain steps match the posterior probability distributions of the parameters.

To illustrate such a scheme, we follow the Metropolis-Hastings algorithm (Hastings 1970), a so-called `walker' is started at a given point in the parameter space, here taken to be the best-fitting values of the parameters found during the initial fit of the model to the data. From this point, a random step is taken in the parameter space with distributions in each parameter drawn from the diagonal of the covariance matrix found during the fit (this is the variance of each parameter). The likelihood of data given the model with these new parameters is then computed. If this is greater than the likelihood at the previous position (i.e. the fit to the data is improved), the walker definitely moves to the new location. If not, the walker moves to the new location, but only with a probability defined by the ratio of the likelihoods at the new and old locations. Therefore, the walker may move or the step may be rejected and the walker may stay at the same location on any iteration. The process is then repeated, with the walker either moving or staying put over a defined number of steps, and the probability distribution of the model parameters given the data is built up as the walker moves.

We use MCMC methods to determine the errors in the model parameters as these can simultaneously and efficiently find the errors in multiple parameters simultaneously and better sample the parameter space than traditional methods, simply stepping through parameter values, performing a fit to the spectrum at each value and evaluating the variation in the goodness-of-fit at each stage. MCMC methods are less vulnerable to local minima in the goodness-of-fit, with their ergodicity meaning that they explore the full parameter space and also less vulnerable to steep gradients in the goodness-of-fit causing the error calculation to simply peg at the hard limits set for the parameter.

Rather than using the Metropolis-Hastings sampler, we use that of Goodman \& Weare (2010), implemented through the {\sc emcee} package in 
{\sc python}\footnote{We use the xspec\_emcee implementation of the python emcee package for X-ray spectral fitting in xspec by Jeremy Sanders 
(http://github.com/jeremysanders/xspec\_emcee)}. The Goodman-Weare algorithm is better able to cope with degeneracies between model 
parameters and is insensitive to the absolute scaling of the probability distributions. We trace multiple walkers through the parameter space simultaneously (for good sampling, the number of walkers should be more than twice the number of model parameters) and the final chain is formed by combining the steps of all the walkers. We simply assume a uniform prior (i.e. make no prior assumptions about the probability distributions of the model parameters except for their hard limits set in the spectral fit).

The first 1000 steps of each walker's chain are discarded to remove bias introduced by the choice of starting location and in order to test that the chains have converged, the calculations are run 320000 times to ensure they converge upon the same parameter values. Finally, in order to ensure that the walkers are adequately sampling the parameter space, the rule-of-thumb is that steps should be rejected less than 75 per cent of the time (i.e. less than 75 per cent of the steps in the chain are simply repeats of the previous step's values). We find that in order to properly sample the parameter space, it is necessary to fit only the combined spectrum from the front-illuminated XIS detectors. Attempting to fit data from the XIS and PIN detectors simultaneously, we find, results in high repeat fractions ($> 99$ per cent) in the Markov chains, most likely caused by the simultaneous fit between the two data sets leading to sharp increases in the likelihood function between parameter values, thus causes the Markov chain to fail to reproduce the probability distribution.

A value for the Galactic column density toward \mrk335\ of
$3.56 \times 10^{20}\pscm$ (Kalberla \et 2005) is adopted in all of the
spectral fits and
abundances are from Anders \& Grevesse (1989).
Luminosities are calculated using a
Hubble constant of $H_0$=$\rm 70\ km\ s^{-1}\ Mpc^{-1}$ and
a standard flat cosmology with $\Omega_{M}$ = 0.3 and $\Omega_\Lambda$ = 0.7.  


\section{Tracking the changes between the 2006 high-flux and 2013 low-flux observations}

\subsection{The spectral characteristics in the high- and low-state}
\label{sect:specdiff}
\begin{figure*}
\begin{center}
\begin{minipage}{0.48\linewidth}
\scalebox{0.32}{\includegraphics[angle=270]{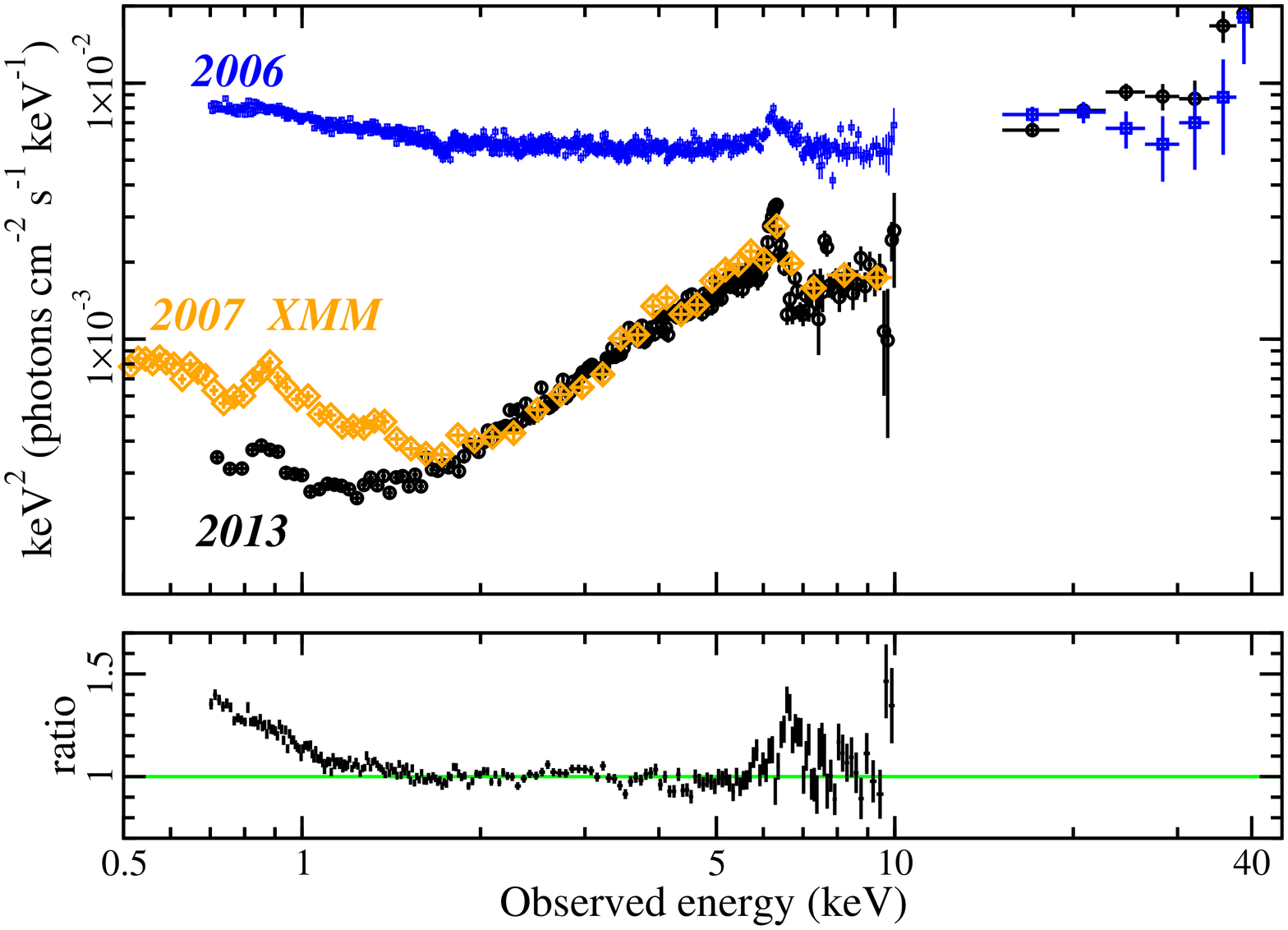}}
\end{minipage}  \hfill
\begin{minipage}{0.48\linewidth}
\scalebox{0.32}{\includegraphics[angle=270]{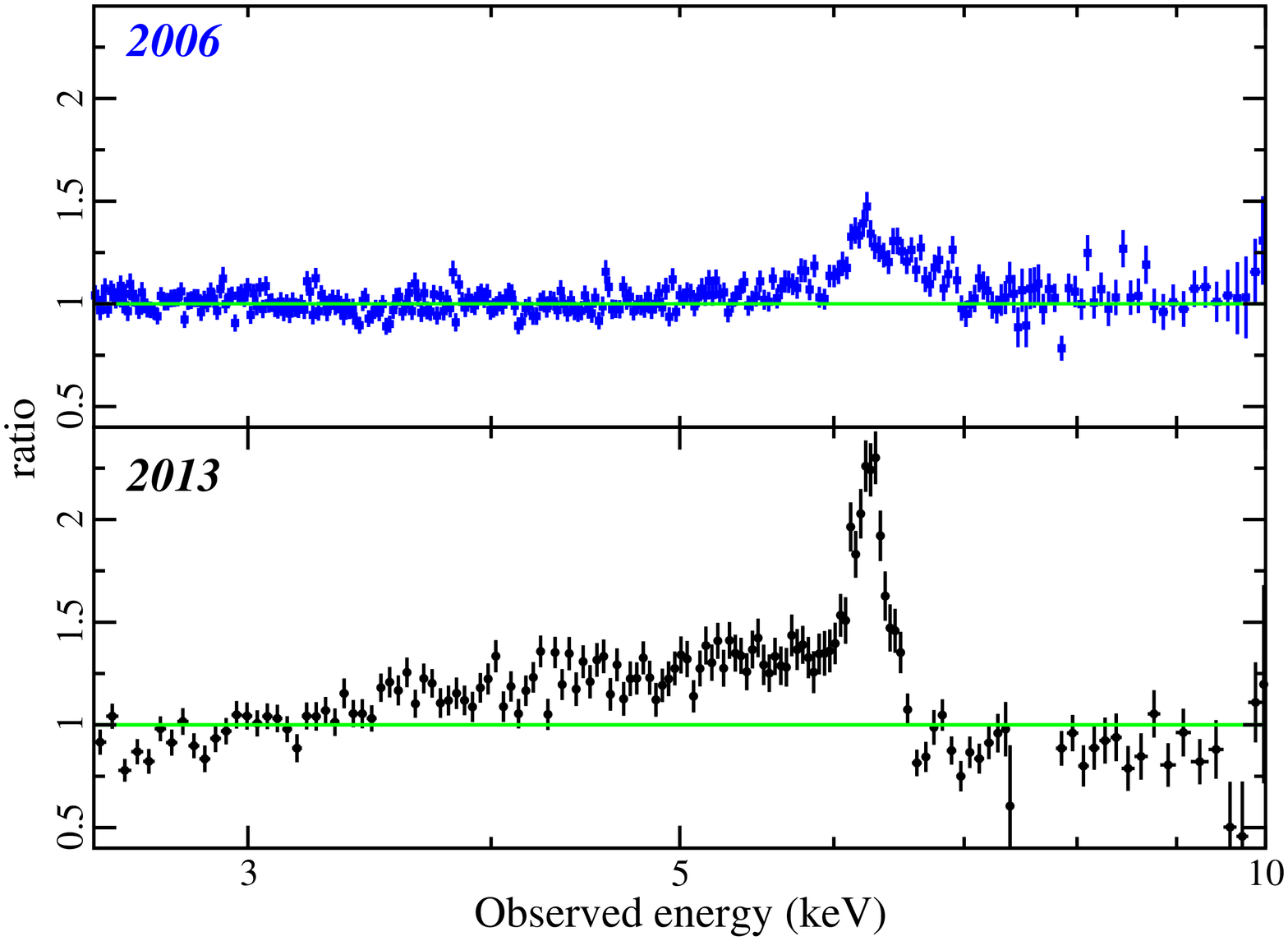}}
\end{minipage}
\end{center}
\caption{Left:  The \suzaku\ spectra of \mrk335\ in a high-flux state in 2006 (blue squares) and in the 2013 low-flux state (black circles) are shown in the top panel.  There is considerable difference below $10\keV$, but the spectra are similar between $15-40\keV$.  The 2007 \xmm\ low-flux state observation is also included for comparison (orange diamonds).  The low-flux spectra are identical above $2\keV$, but the \xmm\ spectrum exhibits excess emission below that energy.  This excess emission could be from the ionised emitter that was present at that time (Grupe \et 2007; Longinotti \et 2008).
The ratio of the difference spectrum (2006 FI data -- 2013 FI data) fitted with a power law ($\Gamma\approx2.3$) between $2-10\keV$ is shown in the lower panel.  Differences are primarily seen above $6\keV$ and below $\sim1.5\keV$.  
Right:  The ratio from a power law fitted between $2.5-4\keV$ and $8-10\keV$ bands is show for the 2006 (top) and 2013 (lower) FI spectra.  In addition to different photon indices and normalisations, significant curvature and sharper features are evident in the low-flux, 2013 observation.  The ratio scale is identical in both panels. }
\label{fig:HighLow}
\end{figure*}

Significant changes between the high-flux state and low-flux state are clearly evident in the panels of Fig.~\ref{fig:HighLow}.  
The average spectrum is much steeper and exhibits fewer distinct features during the 2006 high-flux state.  Strong curvature is seen in the 2013 low-flux observations along with strong, distinct features in the \feka\ band  and around $0.9\keV$.  A power law fitted to the $2.5-10\keV$ band in the low-state renders a much poorer fit ($\redchi\approx 1.7$) than for the high-flux state ($\redchi\approx 1.2$).  

A difference spectrum (the low-flux spectrum is subtracted from the high-flux spectrum) fitted with a power law between $2.5-10\keV$ reveals significant changes in the \feka\ band and below about $2\keV$.  Despite significant changes below $10\keV$ the high-energy spectrum between $15-40\keV$ is nearly identical at both epochs.

\subsection{The lag frequency analysis}
\label{sect:lag}

\begin{figure}
\rotatebox{0}
{\scalebox{0.45}{\includegraphics{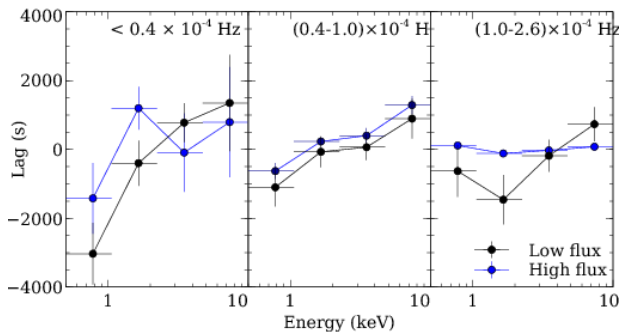}}}
\caption{
The lag as a function of energy at various frequencies during the 2006 (blue data) and 2013 (black data).  Error bars are plotted on all points, but in some cases are smaller than the symbol.
}
\label{fig:lag}
\end{figure}
To obtain additional constraints on the emission components in \mrk335, we study the frequency-resolved time lags. These provide a measure of the time delay between different energy bands (e.g. Zoghbi \et 2010, 2011; Uttley \et 2014). Light curves in four energy bins between 0.5 and 10 keV were produced. The time delay as a function of frequency is calculated between each bin and some reference band, taken in this case to be the whole $0.5-10\keV$ band (excluding the band for which the lag is calculated so the noise remain uncorrelated).  The maximum likelihood method in Zoghbi \et (2013) was used to calculate the lags because of the gaps in the light curves caused by the \suzaku\ low-Earth orbit.

Fig.~\ref{fig:lag} shows the lag-energy plots for three frequency bands for the low- and high-flux observations. In order to directly compare the lag results with that of Kara \et (2013), we use the same two frequency bins. The \suzaku\ observation covers slightly longer time-scales, so we have an extra frequency bin (shown on the left in Fig.~\ref{fig:lag}). Starting with the high flux observations, the lag for the two highest frequencies are very similar to those found in \xmm\ data (Kara \et 2013). On the longest time-scales, a continuum lag characterized by an increase with energy seem to be present, with a possible structure at $\sim1\keV$, though not very significant.

The results for the low flux observation are interesting. The uncertainties in the highest frequencies are large and it is not clear whether there is any significant structure or deviation from a constant. For the middle and left panels (intermediate and lowest frequencies), there is a clear increase in lag with energy (i.e. hard lags). It is intriguing to note that despite the significant difference in the spectrum between the high and low flux observations (Fig.~\ref{fig:HighLow}), the low frequency lags do not seem to change. The low frequency lags are usually interpreted as continuum lags that are intrinsic to the power law spectral component. This is how they were interpreted in the high flux \xmm\ observation (Kara \et 2013).   If the primary component has indeed diminished in the low state as is often suggested in blurred reflection models, the fact we observe the lag in the low-flux data indicates the power law lag is being echoed in another spectral component.


\subsection{Principal component analysis }
\label{section_pca_all}

To investigate the spectral variability between the high and low states in Mrk~335 in a model-independent manner we use principal component analysis (PCA).This is a powerful method for extracting the spectra of multiple spectral components from a variable source (Kendall 1975).
We use the singular value decomposition (SVD, Press \et 1992) based PCA code described in Parker \et (2014a) to compute the principal components. We follow the same method as Parker \et, but we use 100 logarithmic energy bins to show narrow features and longer intervals of 30~ks to compensate for the lower count rate, particularly in the low state. 

\begin{figure}
\includegraphics[width=\linewidth]{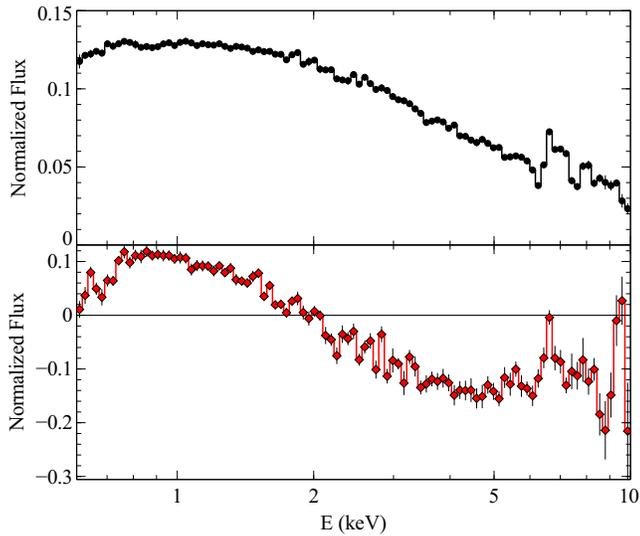}
\caption{First (top, black) and second (bottom, red) principal components returned from PCA of Mrk~335 when all data (both low and high states) is used. }
\label{pcafig_statechange}
\end{figure}

We find two highly significant components in our analysis, with the first component responsible for $\sim98$~per cent of the variability and the second $\sim0.7$~per cent, with the remainder indistinguishable from noise. The spectra of these components are shown in Fig.~\ref{pcafig_statechange}.
The primary component, which corresponds to the bulk of the variability in the state change, shows a general decrease with energy and a narrow iron line at 6.4~keV. Normally, the effect of a narrow emission line is to suppress the spectral variability at that energy (pushing the component towards zero), as distant reflection is relatively invariant. In this case, however, the line appears to be enhanced rather than damped in the primary component. This is probably due to the large interval (7 years) between observations, so some variability can be seen in the iron line - A higher continuum flux should correspond to more distant reflection being observed in general.

\begin{figure}
\includegraphics[width=\linewidth]{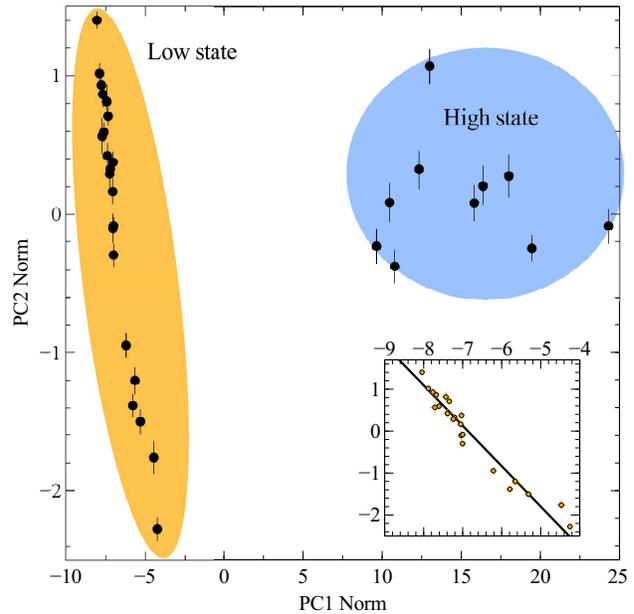}
\caption{Normalizations of the two PCs for each of the 30~ks input spectra. PC1 cleanly divides the data into the two states, and is responsible for the vast majority of the variability (note the different axis scales). In the low state PC2 anitcorrelates strongly with PC1, but there appears to be no such correlation in the high state.  The inset shows a zoom in on the low state points, and a simple linear fit. }
\label{pcafig_pcnorms}
\end{figure}

The normalizations of the two components split the dataset cleanly into the two states. In Fig.~\ref{pcafig_pcnorms} we plot the component normalisations against each other, for each of the input spectra. This helps to shed light on the role of the second component, which is most strongly variable in the low state. This component is clearly anti-correlated with the primary component in the low state, as shown in the inset, and its main role is therefore likely to be as a correction to the primary component, such that it reproduces the dominant variability mechanism in the low state. By subtracting this second component from the first, the low energy variability will be suppressed, and the high energy variability enhanced, making it appear more like the primary component found from PCA of just the low state (see Section~\ref{sect:pca}). The sharp feature around 9~keV appears to be due to noise, and disappears when the data is more highly binned.

We note the difference in the components returned here from those found in NGC~1365 (Parker \et 2014b). NGC~1365 shows very strong evidence of both partial-covering absorption and relativistic reflection(Risaliti \et 2013), however the variability is dominated by changes in the column density  and covering fraction of the absorber (Walton \et 2014). As a result, the components returned from PCA of that source showed a clear and sharp absorption edge at 7~keV. This is also seen in simulations, and in other sources with strong absorption variability (Parker et al., submitted). The absence of such features here suggests that absorption is not the dominant cause of the long-term spectral variability in this source.

\subsection{Blurred reflection fits for the multi-epoch spectral data}
\label{sect:meanref}

The blurred reflection model has been very successful in fitting many Seyferts and NLS1s including \mrk335\ (e.g. Larsson \et 2008; G13; Walton \et 2013).    In this section we examine if the average high- and low-flux spectra can be fitted with self-consistent blurred reflection models.  We consider a simple geometry where the primary X-ray source seen by the observer, modeled as a power law with a high-energy exponential cutoff, also  irradiate the accretion disc.  Some of the emission striking the accretion disc is backscattered into the observers line-of-sight creating a reflection spectrum (e.g. Ross \& Fabian 2005) that is blurred by dynamics in the disc and relativistic effects close to the black hole (e.g. Fabian \et 1989; Laor 1991; Brenneman \& Reynolds 2006).

The photon index and normalisation of the primary power law is free to vary at each epoch while the cutoff energy is fixed at $300\keV$.  Recent \nustar\ observations of \mrk335\ (Parker \et 2014), contemporaneous with the low-flux \suzaku\ observation presented here find no evidence of the cutoff being at $E<200\keV$, therefore fixing it to a high energy well outside the the \suzaku\ PIN band is reasonable.  

The backscattered emission from the disc is modeled with the {\sc reflionx} model in {\sc xspec}.  The iron abundance is free to vary, but is linked between the two flux states.  We consider scenarios in which only the ionisation parameter ($\xi=4\pi F/n$ where $n$ is the hydrogen number density and $F$ is the incident flux) or the model normalisation are allowed to vary at each epoch.  These provide reasonable fits, but better models are obtained when both parameters are free to vary in each spectrum.  This emission component is then blurred for relativistic effects using the {\sc kerrconv} model (Brenneman \& Reynolds 2006).  The dimensionless black hole spin parameter ($a=Jc/GM^2$ where $J$ is the angular momentum of the black hole of mass $M$) and the disc inclination ($i$) are linked between the two epochs as they are not expected to vary.  The disc is assumed to extend down to the inner most stable circular orbit ($\risco$) and the disc emissivity profile is described with a broken power law index ($\epsilon \propto r^{-q}$).  The emissivity profile between $\risco$ and the break radius ($R_b$) is described by $q_{in}$ and beyond  $R_b$ the index is described by $q_{out}$.  Various combinations are attempted to fit the emissivity profile, but the multi-epoch data are most efficiently fit when $q_{out}=3$ and the break radius is fixed at $R_b =6\rg$.

A second {\sc reflionx} component that is not blurred and whose flux remains constant at both epochs is also included to mimic emission from a distant reflector like the torus.  The component would produce the narrow \feka\  emission line at $6.4\keV$ and contribute flux to the $20-40\keV$ band via the Compton hump.  The ionisation parameter of this distant reflector is fixed at $\xi=1\ergcmps$ and the iron abundance is set to the solar value. Adopting a solar iron abundance for the distant torus is not unreasonable without evidence to the contrary.  The iron abundance close to the black hole is often seen to be super-solar in the inner regions of the accretion disc for various possible reasons like the radiative levitation of iron ions in the disc (Reynolds \et 2012) or the effects of returning radiation in the inner disc (Ross, Fabian \& Ballantyne 2002), but the abundances may or may not need be so high at larger distances like in the broad line region or warm absorber (e.g. Komossa \& Mathur  2001; Leighly \& Moore 2004; Fields \et 2005; L13).  The photon index of the power law source illuminating the torus is fixed at $\Gamma=1.9$.  The torus likely sees a different power law photon index than the observer or the inner disc because of different lines-of-sight through the corona.  In addition, given the large light travel time delays the torus is probably not sensitive to  rapid changes in $\Gamma$ that would influence the inner disc. Therefore the photon index of the power law illuminating the torus is fixed to the average photon index observed over long-term monitoring of \mrk335.  These assumptions simplify the fitting, but have little influence on the overall spectral model since the component is secondary to the power law and/or blurred reflector, and since the PIN data are of modest quality.    Note that Walton \et (2013) and Parker \et (2014) treat the distant reflector in \mrk335\ differently than how we have done here, but the results are comparable.

Furthermore, the emission from the torus could have varied over the seven years between observations and there is some evidence to support this in the multi-epoch PCA (Fig.~\ref{pcafig_statechange}).  However, no significant variability was seen in the spectral analyses (e.g. see Table~\ref{tab:hilofit}), likely due to signal-to-noise, so the component is linked for much of the analysis.

The described model produces a reasonable fit to to both spectra, but there are noticeable residuals in the low-flux spectrum.  An excess is seen at approximately $0.9\keV$ that could be well reproduced by the addition of a narrow ($\sigma=1\eV$) Gaussian profile ($\delchi=16$ for 2 additional fit parameters: energy and normalisation).  The $\sim0.9\keV$ feature was not required in the high-flux spectrum with the normalisation being consistent with zero when a Gaussian profile was included.  The emission feature is comparable in flux and energy to that reported by Grupe \et (2007) during the 2007 low-flux state \xmm\ snap-shot observation of \mrk335.

The residuals to this model portray an absorption-like feature at approximately $7\keV$ in the low-flux spectrum that could arise from highly ionised iron in a warm absorber.  Longinotti \et (2013) modeled the complex absorption seen in the intermediate flux-state of \mrk335\ with a triple-zone warm absorber.  These absorbers were then included by G13 when they analyzed the continuum of \mrk335\ during the intermediate flux state.  The hottest absorber in the L13 model would generate features at the observed energies.  The inclusion of the hottest absorber with its parameters fixed to the values found by G13 (and consistent with L13) was a statistical improvement ($\chidof=1.10/3611$), however including all three of the L13 warm absorbers did not enhance the fit further.  Allowing the column density ($\nh$) and ionisation parameter of the hottest absorber to vary did not significantly improve the fit either ($\delchi=4$ for 2 addition fit parameters).  Adding warm absorbers to the high-flux spectrum was unnecessary in agreement with previous studies of \mrk335\ in the high-flux state (e.g. Larsson \et 2008; L13).

\begin{figure}
\rotatebox{270}
{\scalebox{0.32}{\includegraphics{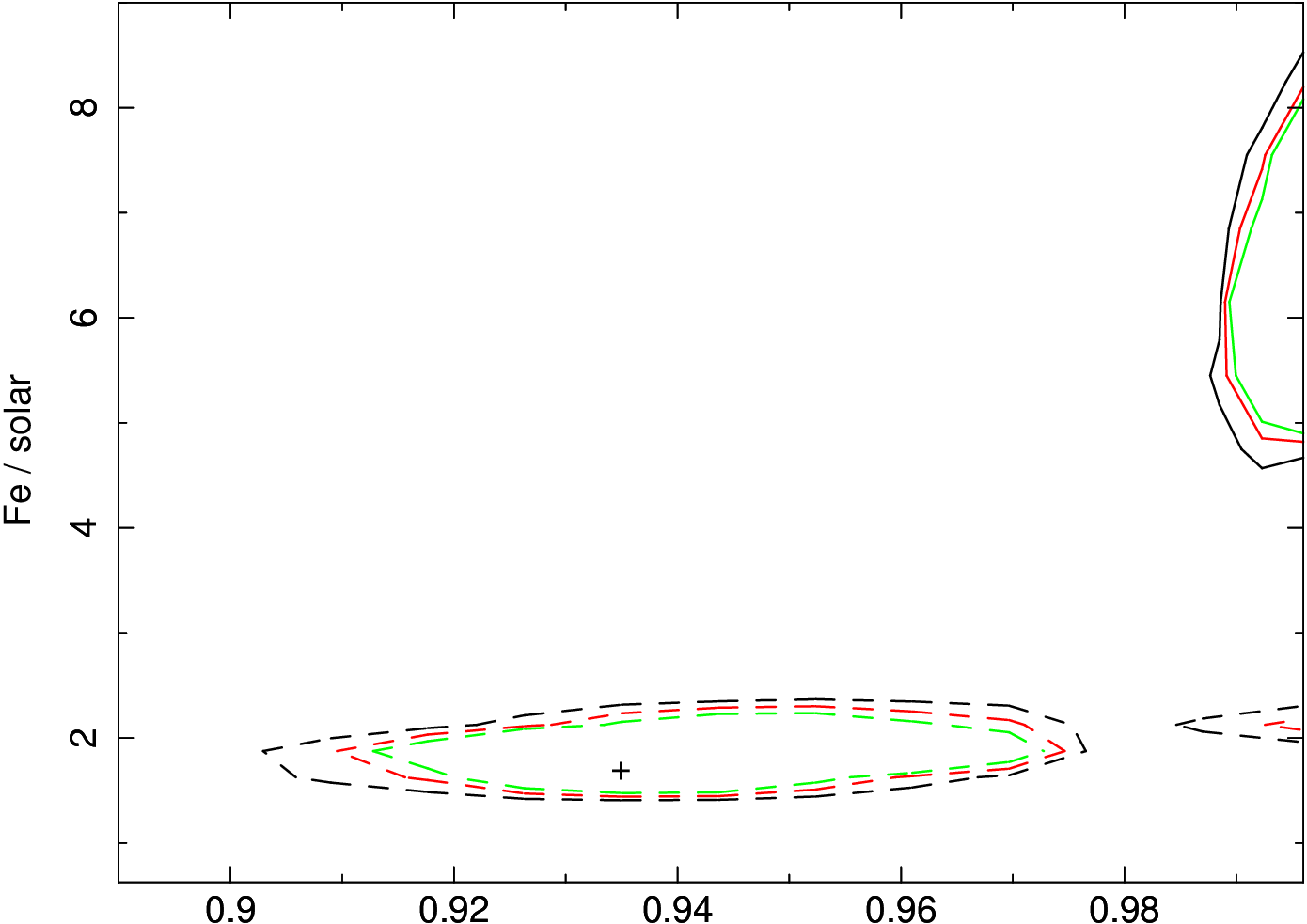}}}
\vspace{5mm}
\caption{
The contour plot showing the dependency between iron abundance and black hole spin parameter during the high-flux (dashed curves) and low-flux (solid curves) observations.  The grid contours are for a delta fit statistic of 2.3, 4.61, and 9.21.  Both fits favour an iron overabundance and high black hole spin, but the best-fit combination differ for both data sets.  
}
\label{fig:cont}
\end{figure}

Analysing the fit in more detail reveals a possible degeneracy between the iron abundance and the spin parameter.  Indeed the error analysis converges on two possible combinations of these parameters.  To examine this further the described reflection model was fitted to the high-flux and low-flux state separately.  The fits to the high- and low-flux states are comparable in all ways (Table~\ref{tab:hilofit}) except that each flux-level is better fit with a different combination of iron abundance and spin parameter.   Fig.~\ref{fig:cont} highlights the discrepancy showing that slightly more extreme values of spin and iron abundance are required in the low-flux spectrum.   Notably, both fits require an iron overabundance compared to solar and the black hole spin that is near maximum ($>0.9$).

It must be emphasized that the degeneracy appears to exist between the iron abundance and spin parameter, and the influence on the determination of other model parameters appears negligible (e.g. compare the parameters in Table~\ref{tab:hilofit} and Table~\ref{tab:Meanfit}).  This mild degeneracy does not adversely influence other spectral analyses, such as the emissivity profile analysis.  Uncertainties in the iron abundance do not affect the accuracy of the measured accretion disc emissivity profiles. An increased iron abundance improves the statistical constraint on the emissivity measured from the iron K$\alpha$ line, with a greater number of photon counts to constrain the contribution from each part of the disc, but since the narrow iron line is convolved with the relativistic blurring kernel, an under- or over-estimate of the iron abundance affects only the overall normalisation of the emissivity profile, not the measured shape of the profile which reveals the geometry of the corona.

\begin{table*}
\caption{The blurred reflection model fitted independently to the high- and low-flux  \mrk335\ spectra.  
The model components and model parameters are listed in Columns 1 and 2, respectively. 
Columns 3 and 4 list the parameter values during the 2006 high state and 2013 low state, respectively.
The superscript $f$ identifies parameters that are fixed.
The inner disc radius ($R_{in}$) is assumed to be fixed at the innermost stable circular orbit.
The reflection fraction ($\mathcal{R}$) is approximated as the ratio of the reflected flux over
the power law flux in the $0.1-100\keV$ band. 
Fluxes are corrected for Galactic absorption and are reported in units of $\ergpscmps$.
}
\centering
\scalebox{1.0}{
\begin{tabular}{ccccc}                
\hline
(1) & (2) & (3) & (4)  \\
 Model Component &  Model Parameter  &  2006 & 2013  \\
                                    &                                   &  High-flux & Low-flux  \\
\hline
Warm Absorber  & $\nh$ ($\times10^{23}$) & & $0.622^f$    \\
           &log$\xi$ [$\erg\cmps$] &  & $3.31^f$   \\
           & $v_{out}$ ($\kmps$) &  & $1800^f$ \\
            \hline
 Incident & $\Gamma$ & $2.18\pm0.01$ & $1.92\pm 0.05$  \\
 Continuum           & $E_{cut}$ (\keV) & $300^f$ & $300^f$ \\
            &$F_{0.5-10 keV}$ & $2.67\pm0.02 \times 10^{-11}$ & $9.67^{+0.76}_{-1.22} \times 10^{-13}$    \\
\hline
  Blurring  & $q_{in}$    & $6.2^{+2.5}_{-1.0}$   & $7.8^{+0.9}_{-1.1}$ \\
             & $a$  & $0.934^{+0.037}_{-0.017}$  & $>0.991$                                      \\
           & $R_b$ ($\rg$)   & $6^f$ & $6^f$ \\
           & $q_{out}$    & $3^{f}$          &  $3^f$ \\
           & $i$ ($\deg$)  & $59^{+7}_{-3}$  &      $58^{+4}_{-6}$                            \\
\hline
  Blurred Reflector  & $\xi$ ($\erg\cmps$)   & $71^{+12}_{-13}$  & $13^{+7}_{-5}$          \\
             & $A_{Fe}$ (Fe/solar)   & $1.9^{+0.3}_{-0.2}$       & $6.7^{+0.8}_{-1.4}$    \\
            &$F_{0.5-10 keV}$ & $6.48^{+3.71}_{-1.22} \times 10^{-12}$ & $2.54^{+1.73}_{-0.78} \times 10^{-12}$    \\
             & $\mathcal{R}$ & $0.33\pm0.02$ & $8.2\pm1.2$   \\
\hline
  Distant Reflector & $\xi$ ($\erg\cmps$)   & $1.0^{f}$  &  $1.0^{f}$        \\
             & $A_{Fe}$ (Fe/solar)   & $1.0^{f}$  &   $1.0^{f}$      \\
     & $\Gamma$ & $1.9^{f}$ & $1.9^{f}$  \\  
            &$F_{0.5-10 keV}$ & $7.73^{+1.22}_{-1.08} \times 10^{-13}$ & $7.84^{+1.08}_{-0.40} \times 10^{-13}$    \\
                \hline
  Additional & $E$ (\eV)    &   &   $886\pm 16$       \\
  Gaussian  & $\sigma$ (\eV)          &  & $1^{f}$     \\
            &$F_{0.5-10 keV}$ & & $1.45^{+0.40}_{-0.70} \times 10^{-14}$    \\
\hline
              Fit Quality & $\chidof$ & $1.09/1895$    & $1.01/1582$ \\
\hline
\label{tab:hilofit}
\end{tabular}
}
\end{table*}

\begin{figure}
\rotatebox{270}
{\scalebox{0.32}{\includegraphics{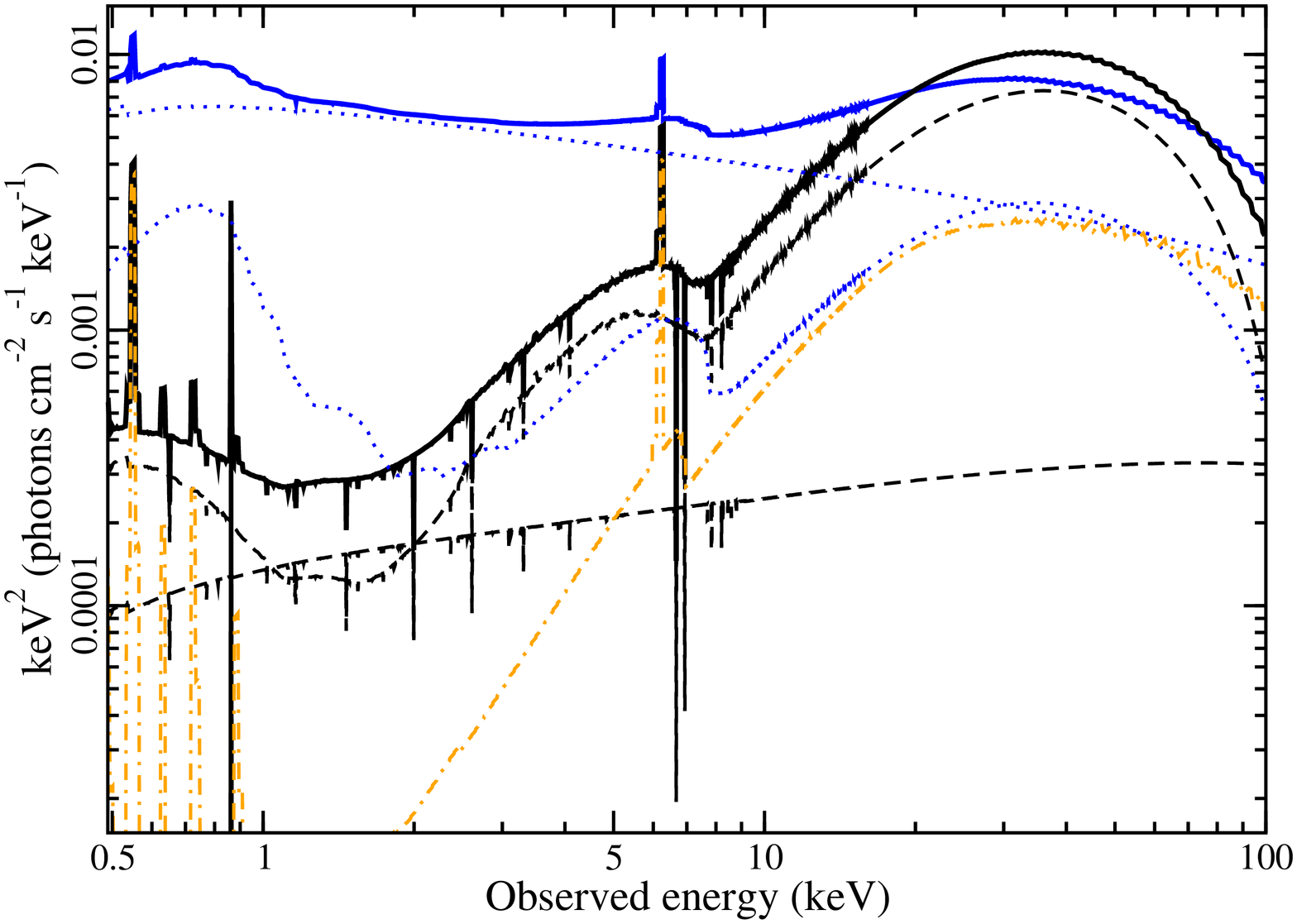}}
\scalebox{0.32}{\includegraphics{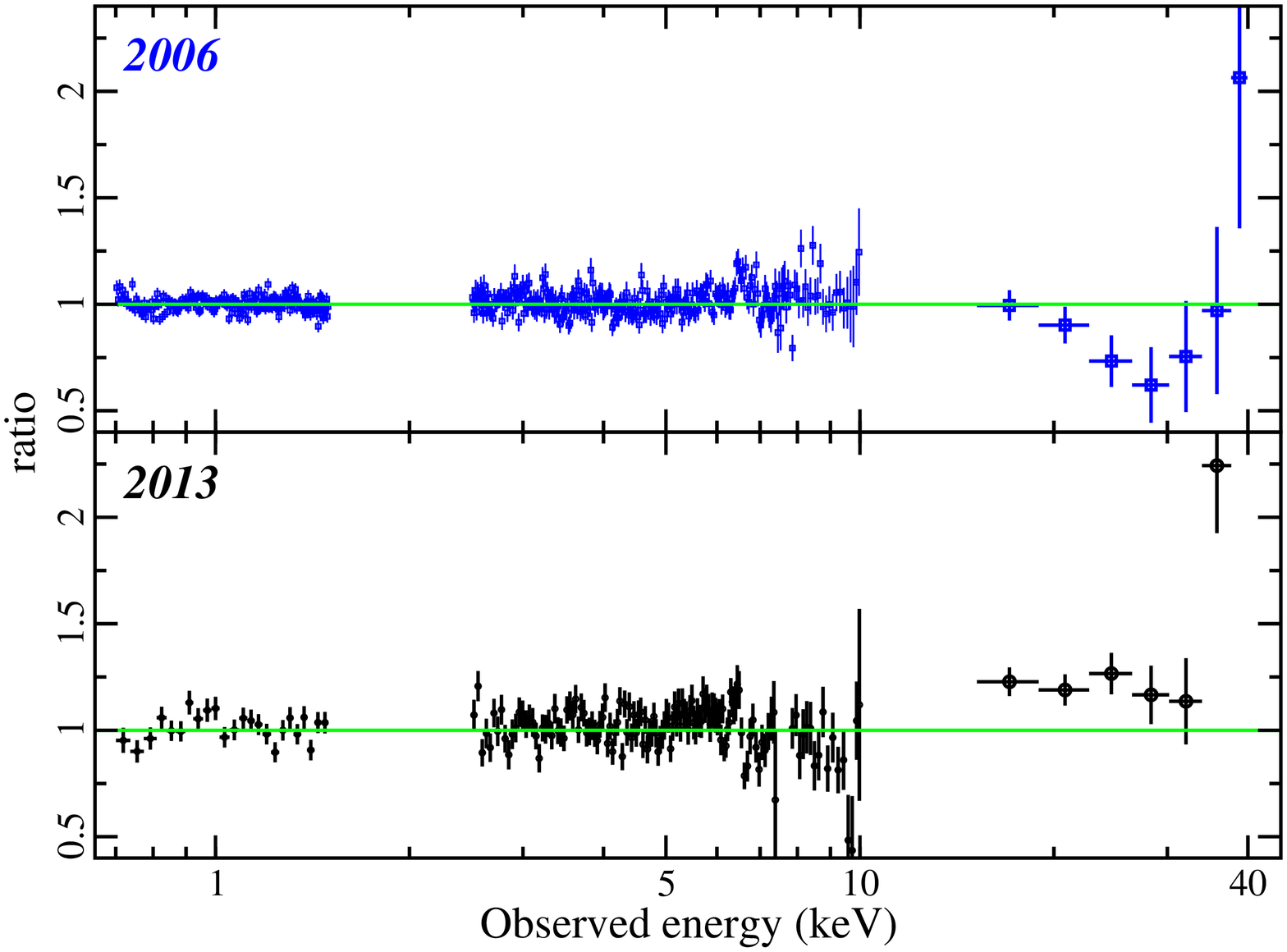}}}
\caption{
Top panel: The blurred reflection model is shown over the $0.5-100\keV$ band.  The blue and black solid curves are the overall 2006 and 2013 models, respectively.  The orange dot-dashed curve is the reflection component from the distant reflector that is identical at both epochs.  The black dashed curves are the reflection and power law component in 2013.  The 2013 spectrum also includes a single warm absorber similar to the hottest absorber in L13.  The spectrum is reflection dominated in the low-flux state and power law dominated in the high-flux state (blue dotted curves).
Lower panel:  The data/model ratio in the high-flux state (top panel) and low-flux state (lower panel).  
}
\label{fig:Refmean}
\end{figure}

Since neither the spin nor iron abundance should vary over these observed time scales, in attempt to achieve a consistent model both epochs were fitted simultaneously while fixing the spin parameter to the different best-fit value found in the low- and high-flux state.  A slightly better fit ($\redchi = 1.10$ versus $1.14$ for the same number of free parameters) was found when the multi-epoch spectra were matched with the higher spin value ($a=0.995$) rather than the lower spin value ($a=0.934$).   This combined fit is presented in Fig.~\ref{fig:Refmean} and Table~\ref{tab:Meanfit}.   

The blurred reflection model is able to describe the average high- and low-flux states of \mrk335\ relatively well.  The well known flattening of the primary continuum with diminishing flux is observed.  At low-flux levels \mrk335\ is reflection dominated with a reflection fraction of $\mathcal{R}=8.2\pm1.2$ (where  $\mathcal{R}$ is the ratio of the reflection and power law component in the $0.1-100\keV$ band).  The high value of  $\mathcal{R}$ is in agreement with the higher spin value of $a=0.995$ according to simulations by Dauser \et (2014).  In comparison, during the bright state  $\mathcal{R}=0.34\pm0.02$.  In accordance, the emissivity profile is steeper in 2013 indicating the X-ray emission is originating from a more compact region and the continuum is more influenced by light bending than in the high-flux state. 

The particularly low $\mathcal{R}$ during the high-flux state suggests that in a lamp post geometry the primary X-ray source is at a large distance from the accretion disc.  The blurring model {\sc kerrconv} is replaced with the {\sc relconv\_lp} (Dauser \et 2013) that calculates an emissivity profile for the disc based on the height of the source above it.   The source height above the disc ($h$) is one of the measured model parameters.  Fitted to the high-flux state of \mrk335, the {\sc relconv\_lp} blurring model generates comparable results and quality of fit as the model in Table~\ref{tab:hilofit} ($\redchi = 1.11$).  The height of the primary X-ray source above the disc is measured to be $h=23^{+6}_{-4} \rg$.  In contrast, applying the same model to the low-flux spectra ($\redchi = 1.17$) results in an upper-limit of $h < 4\rg$ for the height of the X-ray source and is consistent with the values measured in the \nustar\ observation (Parker \et 2014).

\begin{table*}
\caption{The best-fit blurred reflection model fitted simultaneously to the high- and low-flux  \mrk335\ spectra.  
The model components and model parameters are listed in Columns 1 and 2, respectively. 
Columns 3 and 4 list the parameter values during the 2006 high state and 2013 low state, respectively.
The superscript $f$ identifies parameters that are fixed and the superscript $l$ indicates the parameter is linked to the other flux state.
The inner disc radius ($R_{in}$) is assumed to be fixed at the innermost stable circular orbit.
The reflection fraction ($\mathcal{R}$) is approximated as the ratio of the reflected flux over
the power law flux in the $0.1-100\keV$ band. 
Fluxes are corrected for Galactic absorption and are reported in units of $\ergpscmps$.
}
\centering
\scalebox{1.0}{
\begin{tabular}{ccccc}                
\hline
(1) & (2) & (3) & (4)  \\
 Model Component &  Model Parameter  &  2006 & 2013  \\
                                    &                                   &  High-flux & Low-flux  \\
\hline
Warm Absorber  & $\nh$ ($\times10^{23}$) & & $0.622^f$    \\
           &log$\xi$ [$\erg\cmps$] &  & $3.31^f$   \\
           & $v_{out}$ ($\kmps$) &  & $1800^f$ \\
            \hline
 Incident & $\Gamma$ & $2.22\pm0.02$ & $2.05^{+0.03}_{-0.07}$  \\
 Continuum           & $E_{cut}$ (\keV) & $300^f$ & $300^l$ \\
            &$F_{0.5-10 keV}$ & $2.72\pm0.03 \times 10^{-11}$ & $1.02^{+0.02}_{-0.04} \times 10^{-12}$    \\
\hline
  Blurring  & $q_{in}$    & $5.5^{+0.4}_{-0.7}$   & $8.3^{+0.4}_{-0.8}$ \\
             & $a$  & $0.995^f$  & $0.995^l$                                      \\
           & $R_b$ ($\rg$)   & $6^f$ & $6^f$ \\
           & $q_{out}$    & $3^{f}$          &  $3^f$ \\
           & $i$ ($\deg$)  & $60^{+2}_{-3}$  &      $60^l$                            \\
\hline
  Blurred Reflector  & $\xi$ ($\erg\cmps$)   & $56^{+4}_{-6}$  & $3^{+3}_{-2}$          \\
             & $A_{Fe}$ (Fe/solar)   & $4.2^{+0.8}_{-0.3}$       & $4.2^l$    \\
            &$F_{0.5-10 keV}$ & $5.68^{+1.50}_{-0.94} \times 10^{-12}$ & $2.63^{+1.40}_{-0.71} \times 10^{-12}$    \\
             & $\mathcal{R}$ & $0.34\pm0.02$ & $8.2\pm1.2$   \\
\hline
  Distant Reflector & $\xi$ ($\erg\cmps$)   & $1.0^{f}$  &  $1.0^{f}$        \\
             & $A_{Fe}$ (Fe/solar)   & $1.0^{f}$  &   $1.0^{f}$      \\
     & $\Gamma$ & $1.9^{f}$ & $1.9^{f}$  \\  
            &$F_{0.5-10 keV}$ & $8.00^{+0.53}_{-0.38} \times 10^{-13}$ & $8.00^l$    \\
                \hline
  Additional & $E$ (\eV)    &   &   $883\pm 27$       \\
  Gaussian  & $\sigma$ (\eV)          &  & $1^{f}$     \\
                        &$F_{0.5-10 keV}$ &  & $1.46\pm 1.08 \times 10^{-14}$    \\
\hline
              Fit Quality & $\chidof$ & $1.10/3611$    &  \\
\hline
\label{tab:Meanfit}
\end{tabular}
}
\end{table*}

\subsection{The Emissivity Profile of the Reflection Spectrum}
\label{sect:emiss}

Having established that blurred reflection models describe well the high- and low-flux spectra self-consistently, we investigate more closely the relativistically blurred reflection spectrum and the underlying changes that caused the transition from high to low flux by measuring the emissivity profile of the accretion disc; that is the reflected flux as a function of radius on the disc, revealing how the accretion disc is illuminated by the corona.

We employ the method of Wilkins \& Fabian (2011) to compute the emissivity profile, dividing the blurred reflection spectrum into the contributions from successive radii in the disc, described by the \textsc{reflionx} model, convolved with the \textsc{kdblur} blurring kernel in which the inner and outer radius parameters are set accordingly for each annulus and each has a flat emissivity profile. The inclination of the accretion disc (for all annuli) is set to the best fitting value found in the previous fit to the full spectrum, so too are the iron abundance and ionisation parameter of the accretion disc. Also included in the spectral model is the unblurred reflection from distant material and intrinsic absorption by the outflowing material, again with parameters set to the best-fitting values found above. The normalisation (i.e. the contribution) of each annulus to the reflection spectrum is found by minimising $\chi^2$, fitting this model to spectrum over the 3-10\,keV energy range, dominated by the prominent iron K$\alpha$ emission line. In order to constrain the emissivity profile of the inner part of the disc, particularly over the range $5-20\rg$, it is necessary to also fit over the range $3-5\keV$, thereby excluding the core of the line dominated by the outer parts of the disc, $ >20\rg$ (see Wilkins \& Fabian 2011 for a full discussion). Errors are computed by performing a Markov Chain Monte Carlo calculation to obtain the probability distribution of the normalisation of the reflection from each annulus, starting from the best-fitting values found by minimising $\chi^2$.
\begin{figure*}
\begin{center}
\begin{minipage}{0.48\linewidth}
\scalebox{0.40}{\includegraphics[angle=0]{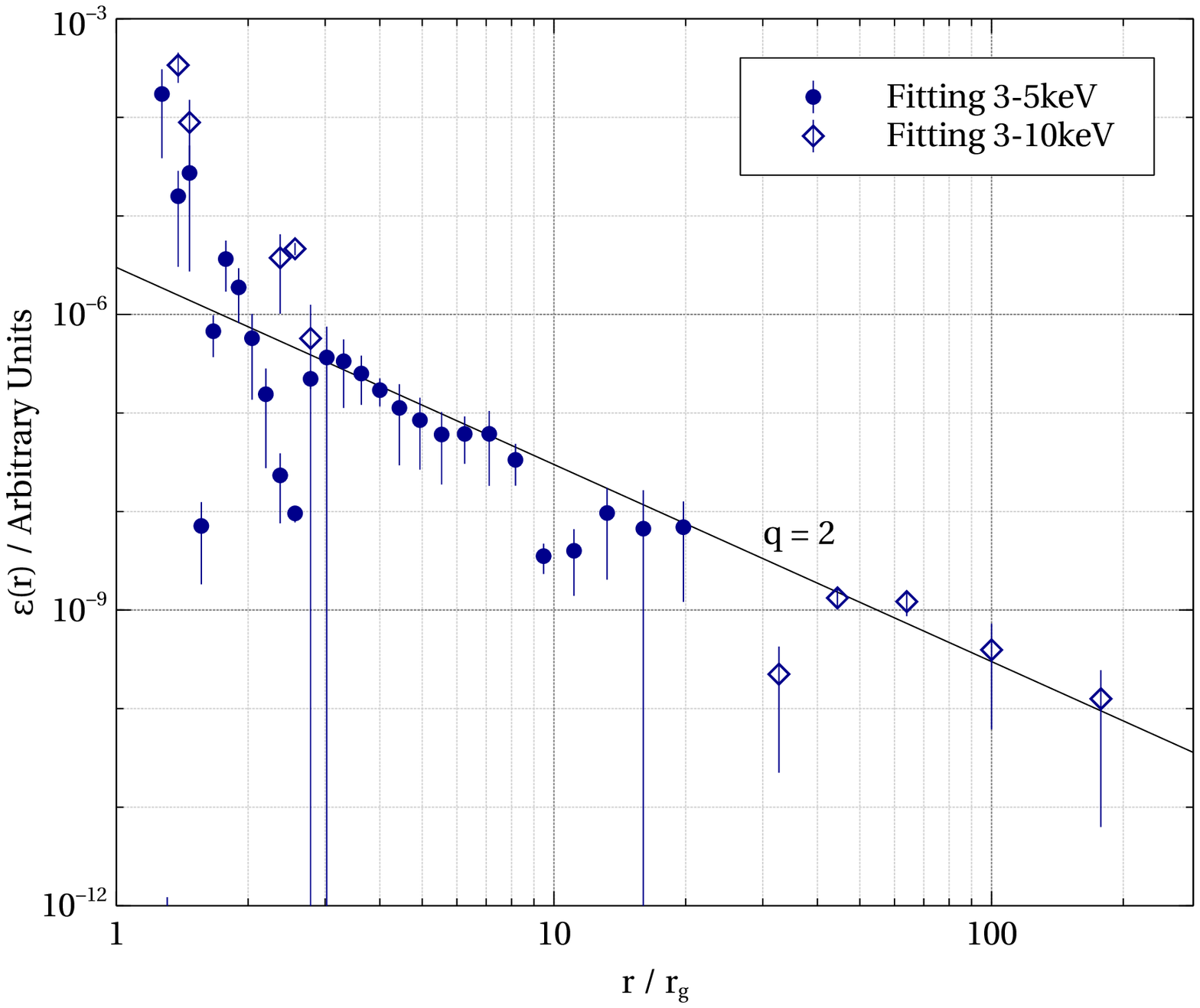}}
\end{minipage}  \hfill
\begin{minipage}{0.48\linewidth}
\scalebox{0.40}{\includegraphics[angle=0]{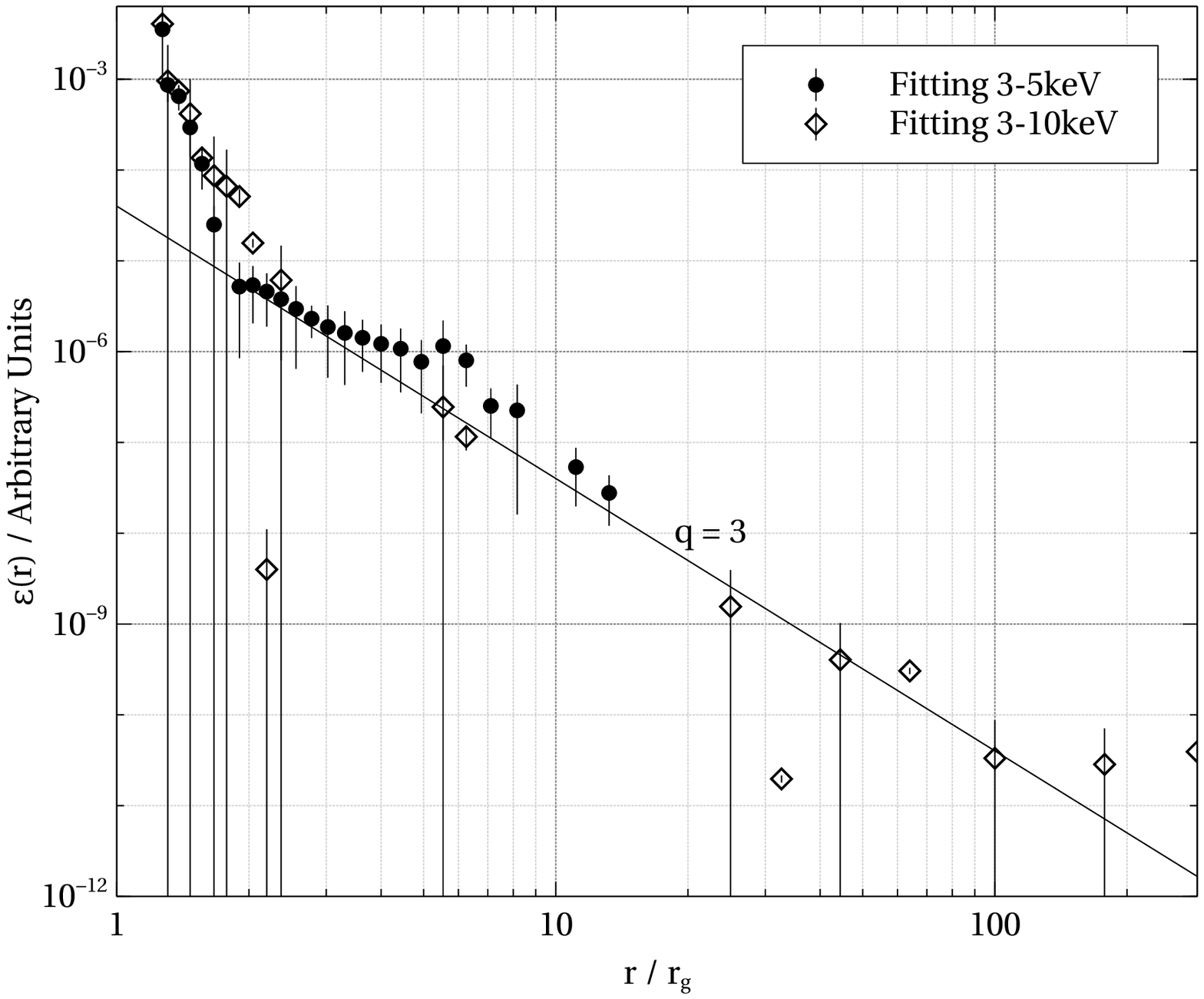}}
\end{minipage}
\end{center}
\caption{The emissivity profiles of the accretion disc in Mrk 335 during the high-flux (left)  and low-flux (right) observations with \suzaku. The emissivity profiles are calculated by decomposing the blurred reflection spectrum into the contributions from successive radii and finding the best-fitting normalisation of the reflection spectrum from each annulus, following the method of Wilkins \& Fabian (2011). Error bars correspond to $1\sigma$ and are computed from Monte Carlo calculations of the probability distribution of each normalisation. }
\label{fig:Emiss}
\end{figure*}

Fig.~\ref{fig:Emiss} shows the measured emissivity profiles of the accretion disc during the high- and low-flux states and it is clear that there is a significant difference in the illumination of the accretion disc by the corona between the two observations. In both cases, the emissivity profile falls off steeply over the innermost part of the disc, with a profile approximating $r^{-9}$ out to $r\sim 3\,r_\mathrm{g}$, as expected for the illumination of the inner parts of an accretion disc around a black hole. The emission reaching the innermost parts of the disc is enhanced by gravitational light bending, focusing rays towards the black hole, as well as time dilation, with observers closer to the black hole measuring more flux from the corona as their time is slowed with respect to those further out, and the emissivity profile is defined in the rest frame of the material in the disc (Wilkins \& Fabian 2012, Miniutti \et 2003, Suebsuwong \et 2006).

During the low-flux state, we see that the emissivity flattens off, with $\epsilon\sim r^{-2}$ to around $r\sim 5\rg$, from where the emissivity profile falls off as $r^{-3}$ over the outer part of the disc. Wilkins \& Fabian 2012 show that such an emissivity profile is expected if the accretion disc is illuminated by a relatively compact corona, confined within $5\rg$ of the black hole.

On the other hand, during the high-flux state, the emissivity profile falls off steeply to $r\sim 3\,r_\mathrm{g}$, from where it simply falls as $r^{-2}$ over the outer part of the disc. The calculations of Wilkins \& Fabian 2012 suggest that such a profile might be expected in the case of a collimated jet-like corona emerging perpendicular to the plane of the accretion disc and extending several tens of gravitational radii. This causes an approximately $r^{-2}$ fall-off in the emissivity profile with a slight steepening at a radius approximately coinciding with the vertical extent of the corona, which can be hard to measure if it occurs far out in the disc (as we would expect for a greatly extended corona).

The measured emissivity profiles of the accretion disc during both the high- and low-flux states show that the drop in flux is associated with a collapse in the X-ray emitting corona from an extended jet-like configuration in the case of \mrk335\ to a confined region within just a few gravitational radii of the black hole. This is consistent with the observations of Fabian \et (2012) who find that the extremely low flux state into which the NLS1 galaxy 1H\,0707$-$495 was observed to drop in 2011 could be explained by a similar collapse in the corona, in this case to within just $2 \rg$ of the black hole and is a more abrupt version of the continuous variation in the extent of the corona as the X-ray flux varies, reported by Wilkins \et (2014).

\subsection{Partial covering models for the multi-epoch spectral data}
\label{sect:meanpc}

An often advocated alternative to the blurred reflection model for \mrk335\ is the partial covering model (e.g. Grupe \et al. 2007, 2008).  In this scenario the observed spectral changes and behaviour arise from changes in a line-of-sight absorber (e.g. Tanaka \et 2004) and do not necessarily require the primary emitter to have altered.  A simple ionised, double partial covering model (e.g. {\sc zxipcf} in {\sc xspec}) resulted in poor fit, but it was useful to highlight important aspects for fitting the model.   To fit the spectra self-consistently the models required a very steep power law continuum ($\Gamma\approx 2.4$) arising from the strong soft excess in \mrk335. The absorbing medium was likely Compton-thick in the low-flux state as it required a column density $\nh > 2\times10^{24}\pscm$.  Thirdly, the narrow \feka\ emission line and excess emission in the PIN band required explanation, and could be consistent with a Compton thick absorber.  The need for a Compton-thick absorber was also supported with the absorber diagnostic diagram of Tatum \et (2013), where the measured hardness ratio (defined as the ratio of the fluxes in the $15-50\keV$ and $2-10\keV$ bands) of $HR\approx 5.5$ in the low-flux state would require about $90$ per cent covering of the primary source by a neutral, Compton-thick medium.  In the high-flux state the covering fraction was negligible ($HR \approx 1$).  With these points in mind a more realistic absorption model was generated.

Rather than modelling the primary continuum with a simple cutoff power law the Comptonisation model {\sc optxagnf} (Done \et 2012) was adopted.  In addition to the  traditional high-temperature, optically-thin corona that generates the hard power law,  {\sc optxagnf} includes emission from a second, cooler Comptonising layer associated with the accretion disc that can imitate the soft excess in AGN.  The inherent curvature in the {\sc optxagnf} model diminishes the need to reproduce the spectral curvature strictly with the ionised absorption alone, thus making spectral fitting easier.  Patrick \et (2011) have previously used a double Comptonisation model ({\sc comptt} + {\sc powerlaw} in {\sc xspec}) to describe the soft excess and hard continuum in \mrk335, so such an approach is not original.

The {\tt optxagnf} model does generate the hard power law emission from the traditional corona, however the normalisation of this parameter is set to zero in this work (specifically the parameter $f_{pl}=0$) and it is replaced with a cutoff power law with $E_{cut}=300\keV$.  The reason is the cutoff in the hard component in {\tt optxagnf} is fixed at  $100\keV$, which would introduce curvature in the PIN spectrum even if \mrk335\ is only detected up to $40\keV$.  The 2006 high-flux spectrum of \mrk335, when the absorption seems negligible, does not seem to require a low-energy cutoff so we assume the same is true at both epochs.   

The model also utilizes the black hole mass ($M_{BH} = 2.6\times10^7 \Msun$) and Eddington luminosity ratio ($L/L_{Edd} = 0.62$), which are adopted from Grier \et (2012) and Gierli\'nski \& Done (2004), respectively.   The distance ($D$) is taken to be $103\Mpc$.   The coronal radius ($R_{cor}$) and outer disc radius ($R_{out}$) are also kept fixed during data modelling as allowing them to vary did not enhance the fits.  The black hole spin parameter was not well constrained when left free to vary, but it did significantly influence the fit quality.  As such three different fits were attempted with the spin parameter fixed at $a=0$, $0.5$, and $0.998$.  The model with a non-rotating black hole was a significantly better fit to the data than the other models invoking higher spin, hence the parameter was set to $a=0$.

\begin{figure}
\rotatebox{270}
{\scalebox{0.32}{\includegraphics[angle=90]{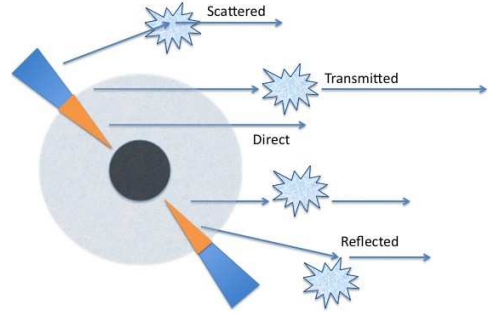}}}
\caption{
Depiction of the partial covering scenario proposed for \mrk335.  The primary X-ray emission arises from the traditional corona shown as a halo around the black hole and second Comptonising component attributed to the inner disc (shown in orange) that reproduced the soft excess shape in the spectrum.  Optically thick clouds in the line of sight obscure some of the primary emission, but as the source is not entirely covered there is a direct (unabsorbed) component that reaches the observer.   Clouds out of the line of sight contribute to the scattered and reflected emission that can reproduce the Fe~K emission lines and Compton hump.  The primary difference between the two flux states present here is there are no obscuring clouds in the line-of-sight during the high-flux state in 2006 (i.e. the covering fraction is $0$) whereas the covering fraction is $\sim96$ per cent in the low-flux state (see Table~\ref{tab:myTmeanfits}).
}
\label{fig:myTpic}
\end{figure}

The requirement of potentially Compton-thick absorption in the low-flux state motivated the adaptation of the {\sc mytorus} model (Murphy \& Yaqoob 2009) to describe the absorption and reprocessing of the partial covering absorber in a self-consistent manner.  Specifically, the {\it decoupled mode} example described by Yaqoob (2012) for NGC~4945 is adopted to mimic a patchy absorber covering the primary X-ray source.  The principal difference between the Yaqoob (2012) model for NGC~4945 and that proposed here for \mrk335\ is that the X-ray emitting region is larger than the obscuring clouds and thus can be partially covered (see Fig.~\ref{fig:myTpic} compared to figure 2 of Yaqoob 2012).  In this scenario the non-variable primary source (modelled with {\sc optxagnf}) is surrounded by a clumpy distribution of absorbers and the observed spectrum is generated from three different processes that occur in this clumpy medium.   The model includes: (i) emission that transverses material in the line-of-sight (transmitted emission); (ii) reflection of the intrinsic continuum by material out of the line-of-sight; and (iii) scattering emission from material in the line-of-sight.  The two latter components are each associated with their own fluorescent line emission (\feka\ and \fekb) and Compton hump emission at $E>10\keV$, thus can offer a natural explanation for the spectral features in \mrk335.  For simplicity, the column densities of each absorber are linked, but could be allowed to vary to reproduce more complex medium.  The specific {\sc mytorus} table used is for a plasma temperature of $50\keV$.  In addition, we include a direct component that reaches the observer without interacting with any obscuring clouds (Fig.~\ref{fig:myTpic}).  
\begin{figure}
\rotatebox{270}
{\scalebox{0.32}{\includegraphics{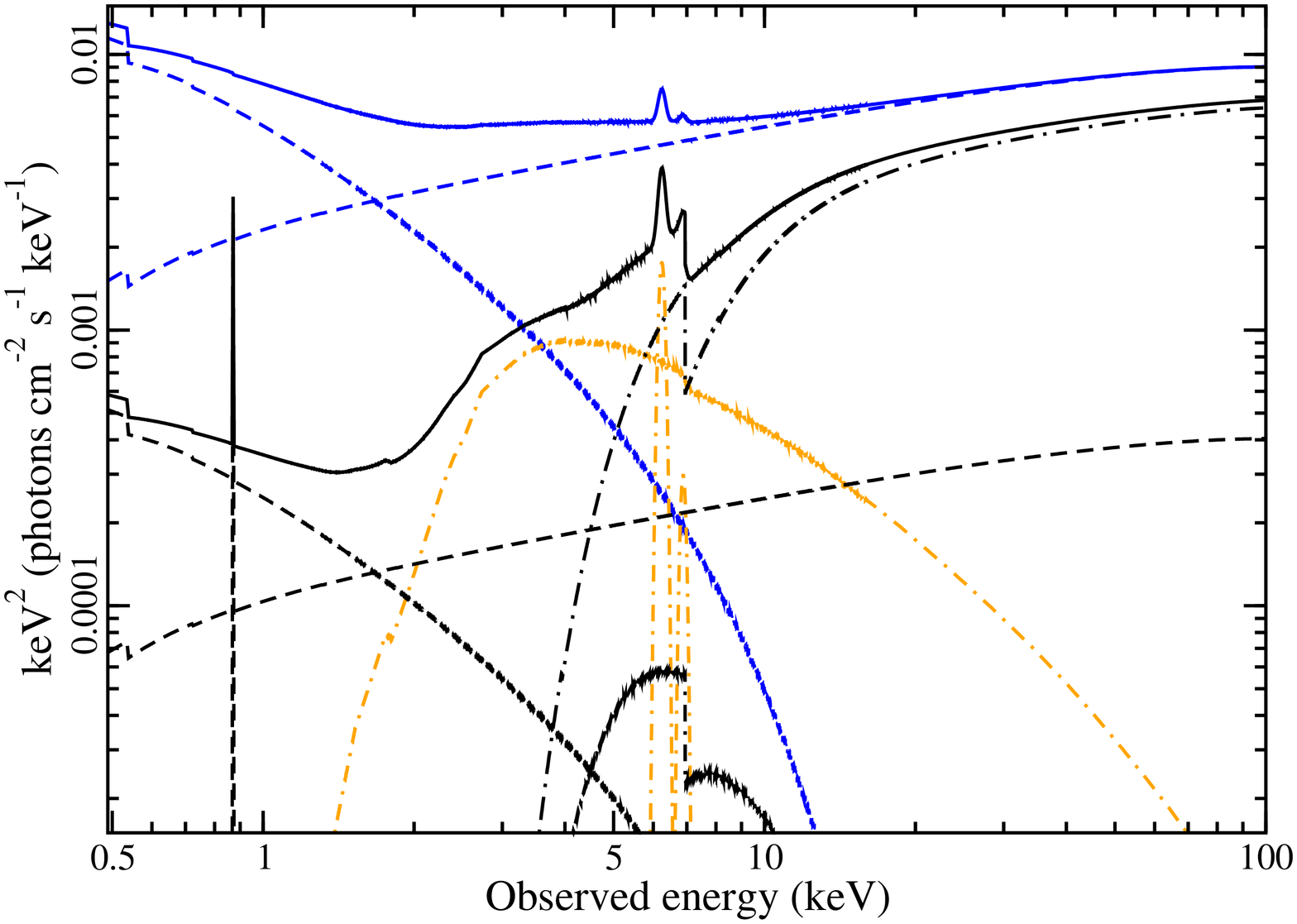}}
\scalebox{0.32}{\includegraphics{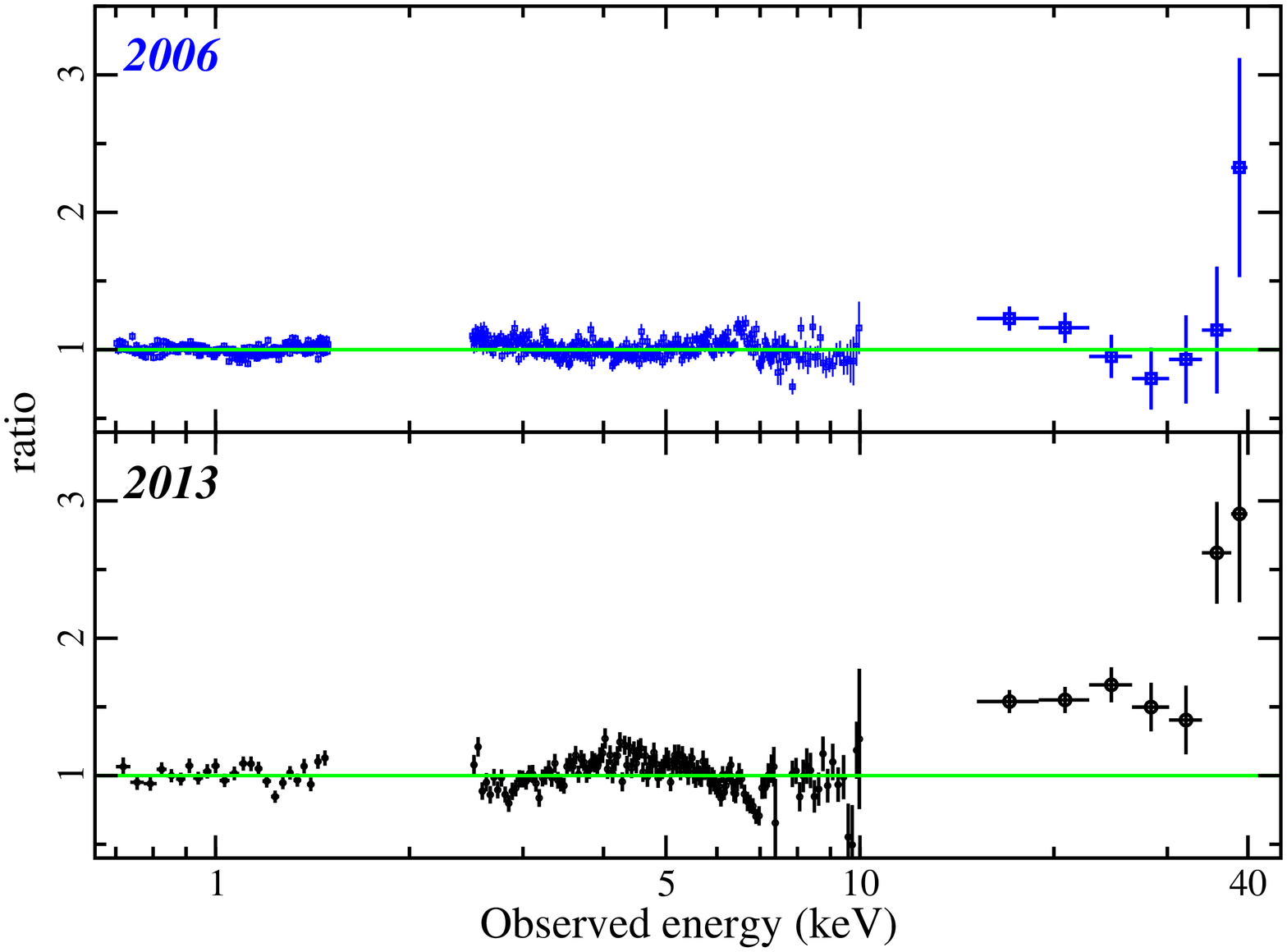}}}
\caption{
Top panel: The partial covering model is shown over the $0.5-100\keV$ band.  The blue and black solid curves are the overall 2006 and 2013 models, respectively.  The high-flux state is completely unabsorbed and the direct Comptonisation continuum is shown as the blue dashed curves.  There is scattered emission that is responsible for the narrow Fe~K line emission and some emission above $\sim10\keV$.  This component is comparable at both epochs (orange dot-dashed curve).  The fluorescent emission originating from the reflected component is modelled, but is negligible in both spectra.  The black dot-dashed curve is the Comptonisation continuum that is absorbed in the line-of-sight along with some ($\sim 5$ per cent) unabsorbed primary emission (black dashed curves).   An emission line at $E\sim 0.9\keV$ is included in the low-flux state.  There is also a warm absorber included in the low-flux spectrum, but it is not shown for reasons of clarity.   
Lower panel:  The data/model ratio in the high-flux state (upper) and low-flux state (lower). 
}
\label{fig:myTmean}
\end{figure}
\begin{table*}
\caption{The best-fit absorption model for the high- and low-flux  \mrk335\ spectra.  
The model components and model parameters are listed in Columns 1 and 2, respectively. 
Columns 3 and 4 list the parameter values during the 2006 high state and 2013 low state, respectively.
The superscript $f$ identifies parameters that are fixed and the superscript $l$ indicates the parameter is linked to the other flux state.
The scattering and reflecting absorbers each have corresponding line emission component.  These are not listed specifically since their parameters are identical to their corresponding absorber.  $A$ is the scaling factor and $C_f$ is the covering fraction.  }
\centering
\scalebox{1.0}{
\begin{tabular}{ccccc}                
\hline
(1) & (2) & (3) & (4)  \\
 Model Component &  Model Parameter  &  2006 & 2013  \\
                                    &                                   &  High-flux & Low-flux  \\
\hline
 Warm Absorber  & $\nh$ ($\times10^{23}$) & & $6.39\pm4.29$    \\
           &log$\xi$ [$\erg\cmps$] & & $3.13\pm0.11$   \\
           & $v_{out}$ ($\kmps$) & & $3060^{+0}_{-3060}$ \\
            \hline
 Incident & $\Gamma$ & $1.62\pm0.09$ & $1.62^{l}$  \\
 Continuum           & $E_{cut}$ (\keV) & $300^f$ & $300^l$ \\
                        & $kT$ (\keV) & $1.97\pm0.72$ & $1.97^l$ \\
                     & $\tau$          & $3.75\pm0.88$ & $3.75^l$ \\
                     & $M_{BH}$ ($\Msun$) & $2.6\times10^{7}$$^{f}$ & $2.6\times10^{7}$$^{l}$ \\
                     & log$L/L_{Edd}$ & $-0.27^f$ &$-0.27^l$  \\
		   & $D$ (\Mpc) & $103^f$ &$103^l$  \\
                     & $a$             & $0^{f}$   & $0^{l}$ \\
                     & $R_{cor}$ ($\rg$) & $10^f$ & $10^l$ \\
                     & log($R_{out}$) [$\rg$] & $5^f$ & $5^l$ \\
            &$F_{0.5-10 keV}$   & $6.74\pm0.03 $ & $6.74^l$    \\
                        &($ \times 10^{-11} \ergpscmps$)  &  &     \\
\hline
  Line-of-sight  & $\nh$ ($\times10^{23}$)    & $5.74^l$ & $5.74\pm0.04$    \\
  Absorber  & $C_{f}$    & $0^{f}$        &         $0.96\pm0.01$   &  \\
\hline
  Scattering  & $\nh$$_{S90}$ ($\times10^{23}$)    & $0.69^l$ & $0.69\pm0.07$          \\
  Absorber  & $\tau_{S90}$          & $0.104^{l}$ & $0.104^{+0}_{-0.003}$     \\
                &  $A_{S90}$  &  $810^l$  &  $810\pm348$     \\
\hline
  Reflecting & $\nh$$_{S0}$ ($\times10^{23}$)    & $\nh$$_{S90}$  &   $\nh$$_{S90}$       \\
  Absorber  & $\tau_{S0}$          & $\tau_{S90}$ & $\tau_{S90}$     \\
                &  $A_{S0}$  &  $69^l$  &  $<69$     \\
                \hline
  Additional & $E$ (\eV)    &   &   $890\pm20$       \\
  Gaussian  & $\sigma$ (\eV)          &  & $1^{f}$     \\
                        &$F_{0.5-10 keV}$ &  & $1.80^{+0.51}_{-0.56} \times 10^{-14}$    \\
          \hline
              Fit Quality & $\chidof$ & $1.22/3612$     \\
\hline
\label{tab:myTmeanfits}
\end{tabular}
}
\end{table*}

The model can approximately describe the differences between the two epochs as arising from changes in the covering fraction of the direct line-of-sight absorber  ($\chidof=1.31/3616$).  The scattered and reflected components are constant between the two epochs and the reflected component is negligible compared to the scattered emission.  For the high-flux state the model reasonably accounts for the iron line emission and PIN emission.  However, as with the blurred reflection model, residuals remain in the low-flux state.  The addition of a narrow ($\sigma=1\eV$) emission line at $E\sim0.9\keV$ in the low state is a significant improvement ($\delchi=86$ for the 2 additional fit parameters energy and normalisation).  The absorption-like feature at about $7\keV$ appears deeper in this model.  The hottest absorber used in L13 and G13 was introduced to fit the high energy residuals.  Allowing the parameter of this component to vary produced a significantly improved fit ($\delchi=246$ for the 3 additional fit parameters column density, ionisation and redshift).  Introducing all three warm absorbers used by L13 did not significantly improve the fit further  even if all parameters were allowed to vary freely.

The picture described reproduces the high-flux state very well, but not the low-flux spectrum ($\chidof=1.22/3612$) as significant excess residuals are seen between $3-6\keV$ and $10-40\keV$.  The similarity in the high- and low-flux spectra above $10\keV$ and similar flux of the narrow iron line suggests that the primary continuum and emitters of the scattered/reflected components do not vary significantly between the two epochs.  The line-of-sight absorption is very different as $96$ per cent of the primary continuum is absorbed by nearly Compton-thick material in the low-flux state, where as the high-flux state is unobscured. The final model and fit parameters are presented in Fig~\ref{fig:myTmean} and Table~\ref{tab:myTmeanfits}.

The partial covering fit can certainly be improved in a number of ways.   For example, adding two more partial covering absorbers, one that is ionised and one that is neutral, improves the fit quality significantly ($\delchi = 222$ for 5 additional free parameters and a $\redchi=1.16$).  Still more absorbers could be added to improve the fit quality further, but degeneracies are an obvious concern.  While any absorber in the vicinity of the black hole is likely to be complex and exhibit density, temperature, and ionisation gradients it becomes impossible to justify adding more components without evidence like that which may eventually become available with future calorimeter data.

The current partial covering model adopts a rigid assumption that the X-ray continuum (power law and soft Comptonisation) is non-variable over the seven years between observations.  Allowing the continuum components to differ in the high- and low-flux states significantly improves the overall partial covering fit ($\delchi = 302$ for 4 additional free parameters and a $\redchi=1.13$).  The differences in the soft Comptonisation component are negligible between the two flux levels.  The temperature and optical depth are approximately $kT \approx 1.4\keV$ and $\tau \approx 4.5$, respectively at both epochs.  This is predictable when viewing residuals in Fig.~\ref{fig:myTmean} that show the spectra are well fit with the same continuum model below $\sim2\keV$.  The difficulty for the model arises in simultaneously addressing the $2-6\keV$ curvature and the high flux above $15\keV$.  To achieve this the model requires an inverted power law ($\Gamma=0.7\pm0.1$) in the low-flux state.  Such spectra have not been reported before for radio-quiet AGN and seem unphysical.

\subsection{Partial covering of the blurred reflection spectrum}
\label{sect:pcref}

There is the alternative possibility that when bright the X-ray emitting region in \mrk335\ is dominated by compact coronal emission and blurred reflection, however the excursions to low-flux levels is driven by partial covering absorption.  
Such hybrid models are rather complex and push the limits of what can be reasonable discerned from current models and data quality.

The reflection spectrum for \mrk335\ in the 2006 high state (Table~\ref{tab:Meanfit}) is adopted as the intrinsic continuum for the low-flux state in 2013.
Applying the partial covering model (and warm absorber), as in Section~\ref{sect:meanpc}, while preserving the continuum did not result in a good fit to the low-flux spectra ($\redchi=1.42$).  As discussed in the previous section, maintaining the continuum constant over 7~years is a rigid demand.  Therefore, the data are refit while allowing the power law continuum parameters (normalisation and photon index) and the reflection parameters (ionisation parameter and normalisation) to vary as well as the absorbers.  The blurring parameters are initially kept fixed to the average values.

The model results in a reasonable fit of the data ($\chidof = 1.12/3609$).  In this scenario, the low-flux state is achieved through considerable variability of continuum components that are significantly covered ($\approx 96$ per cent) by a Compton-thick medium ($\nh \approx 6\times10^{24} \pscm$).  As with other models, the challenge is once again to simultaneously describe the curvature between $2-6\keV$  and the high flux in the PIN band during the low-flux state.  To accommodate, the power law flattens substantially from $\Gamma \approx 2.2$ in the high state to $\Gamma \approx 1.5$ in the low state.  As previously mentioned, the spectrum seems particularly hard for a radio-quiet AGN.   However, the flatter intrinsic spectrum has the net effect of increasing the flux across the band, and also flattening the reflection component.  Consequently, the spectrum at high energy is dominated by the blurred reflection component  and in turn the reflection fraction is high ($R\approx 3.5$ compared to $0.32$ in the high state), rendering the spectrum reflection dominated.  

Partial covering of a blurred reflection spectrum can account for the low state in \mrk335, but only with significant changes to the X-ray emitting components, which may be unrealistic.  In such a scenario the corona would need to collapse to a few gravitational radii to reproduce the reflection dominated spectrum.  In order to partially cover the compact source requires the Compton-thick absorber to be confined to within a similarly small region, which is challenging (Reynolds 2012).

\section{Detailed examination of the low-flux state} 

The 2013 \suzaku\ pointing of \mrk335\ marks the longest exposure of the AGN in its low-flux state and the variability in these data are examined in detail here.   Given the difficulties in explaining the multi-epoch data  and the low-state with partial covering in this (Section~\ref{sect:meanpc}) and previous works (e.g. G12), concentration will be placed on discussing the intra-observation variability in the low state in terms of the blurred reflection model.

\subsection{Rapid variability} 
\label{sect:rvar}
\begin{figure}
\rotatebox{270}
{\scalebox{0.32}{\includegraphics{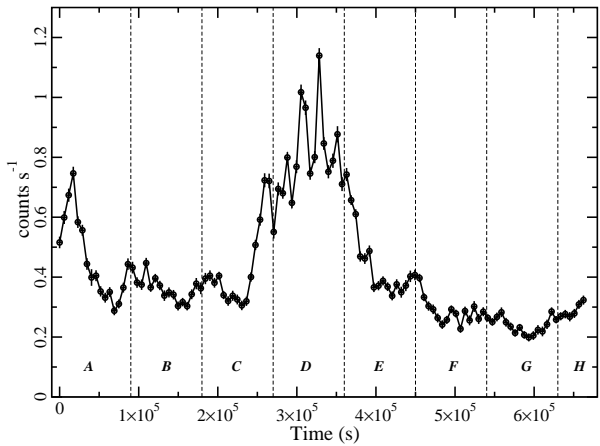}}}
\caption{
Top panel: The $0.6-10\keV$ XIS light curve of \mrk335\  during the low-flux state observation in 2013.  The data are binned on orbital time scales ($5760\s$).  The black line connecting the data points is simply to guide the reader.  The vertical dashed lines mark time segments that will be analysed in Section~\ref{sec:trsv}.  
}
\label{fig:lc}
\end{figure}

The $0.6-10\keV$ light curve of \mrk335\ is displayed in Fig.~\ref{fig:lc} with the data in orbital bins ($5760\s$).  The broadband light curve exhibits significant variability over the observation with the most distinguishing event being a ``flare'' approximately $300\ks$ from the start of the observation.  The flux more than doubles during the brightening incident that lasts about $90\ks$ before returning to pre-flare brightness.   Given the duration of the occurrence it is difficult to draw direct analogies with the very rapid  flares seen in AGN that last only several kilo-seconds.

\begin{figure}
\rotatebox{270}
{\scalebox{0.32}{\includegraphics{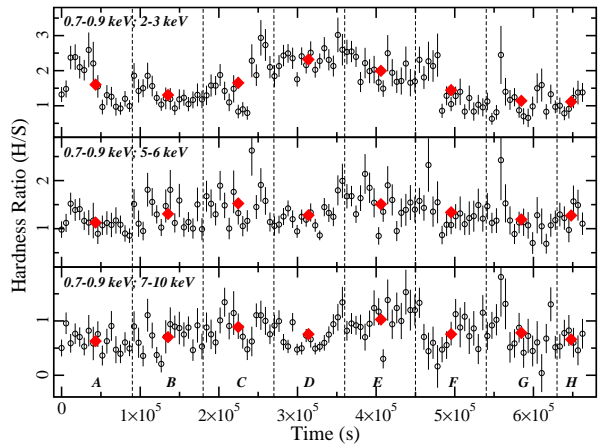}}}
\caption{
The hardness ratio ($H/S$) plotted in various energy bands (shown in each panel) as a function of time.  The vertical dashed lines delineate the same time segments shown in Fig.~\ref{fig:lc}.  The red diamonds are the average HR value in each time segment.  The standard deviation on these points is plotted, but they are smaller than the symbol.  Spectral variability is negligible between the high and low energy bands (bottom two panels), but there does appear to be spectral hardening during the flare when the soft and intermediate energy bands are compared (top panel).}
\label{fig:hr}
\end{figure}

The flare is present in all energy bands and in examining the spectral variability as a function of time (Fig.~\ref{fig:hr}) it appears the hardness ratios between low-energy ($<1\keV$) and high-energy bands ($>3\keV$) are approximately constant.  However the hardness ratio between the soft and intermediate ($2-3\keV$) band does exhibit hardening during the flare that endures for approximately $90\ks$ after the brightness diminishes.
\begin{figure}
\rotatebox{270}
{\scalebox{0.32}{\includegraphics{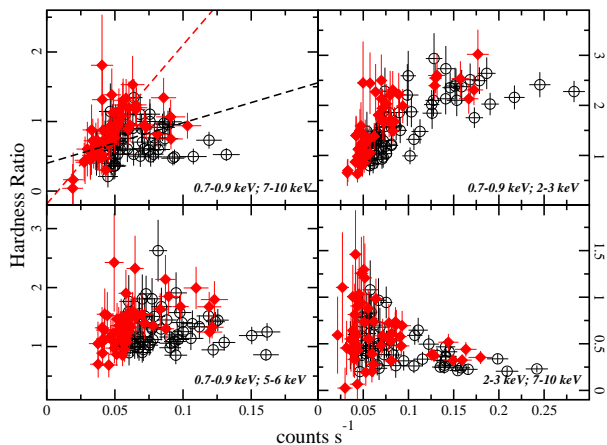}}}
\caption{
Four different hardness ratios are plotted against corresponding count rates.  Data from before and after the flare are displayed in black open circles and red diamonds, respectively.   The top left panel shows that the $HR$ and count rate are correlated after the flare, but not before the flare.
}
\label{fig:hrcr}
\end{figure}

Investigating the hardness ratio as a function of count rate before and after the flare revealed interesting behaviour in one particular comparison.
The spectral variability between the lowest and highest energy bands appeared to become correlated with the count rate after the x-ray flare.  Fig.~\ref{fig:hrcr} (top left panel) shows that prior to the flare (time segments A, B, and C; black circles) there was no significant correlation between the count rate and hardness ratio.  However, after the flare (calculate for time segments E, F, G, and H; red diamonds) the spectrum hardens with increasing count rate.  In addition, both these energy bands are reflection dominated if the presented reflection scenario is correct.  

Similar behaviour was discovered in the NLS1 I~Zw~1 (Gallo \et 2007), but a clear origin was evasive.  The behaviour is often clearly seen in stellar mass black holes that exhibit proper ``state changes'', but the time scales at which these changes occur in AGN do not correspond with similar time scales in stellar mass black holes.

\begin{figure}
\rotatebox{270}
{\scalebox{0.32}{\includegraphics{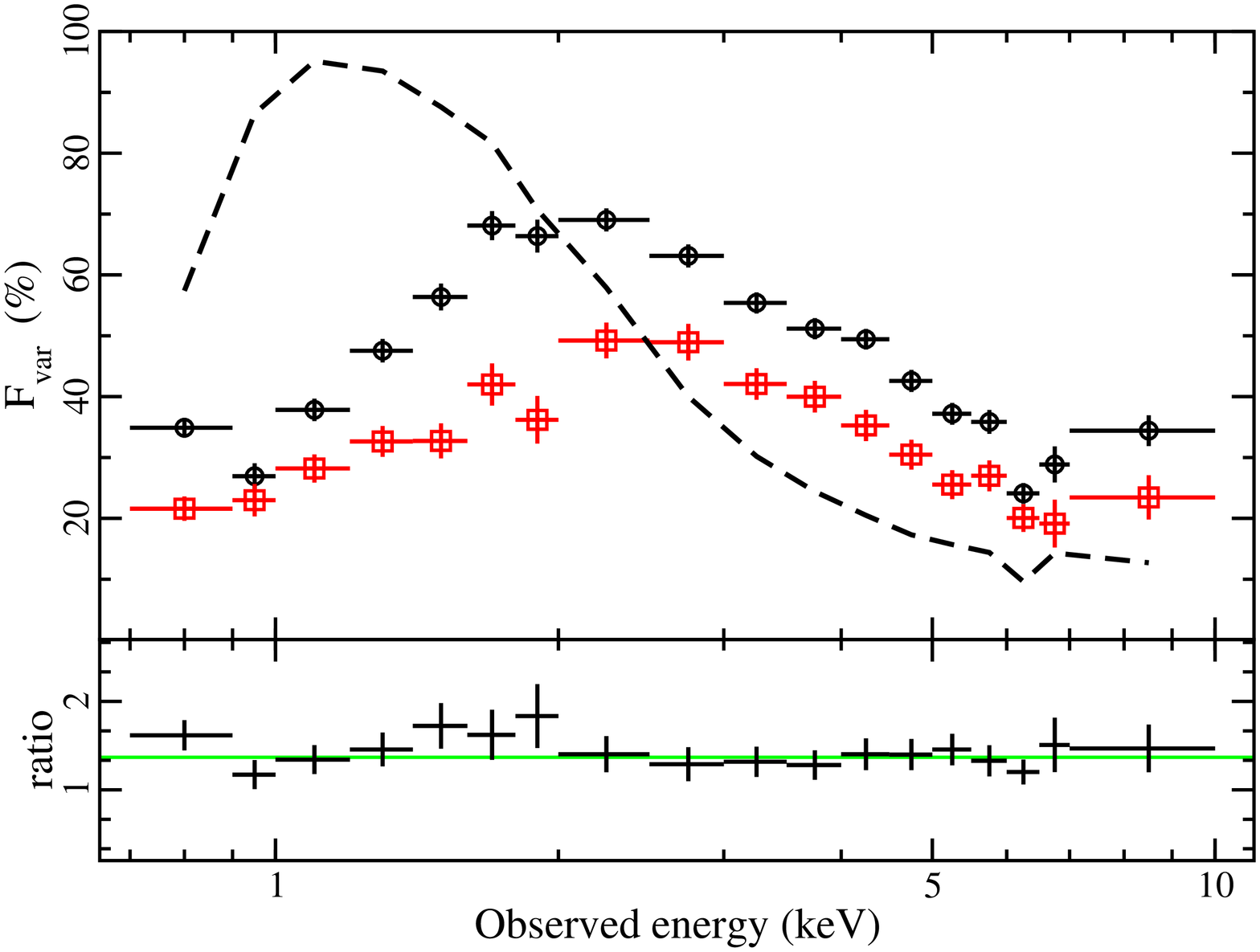}}
\scalebox{0.32}{\includegraphics{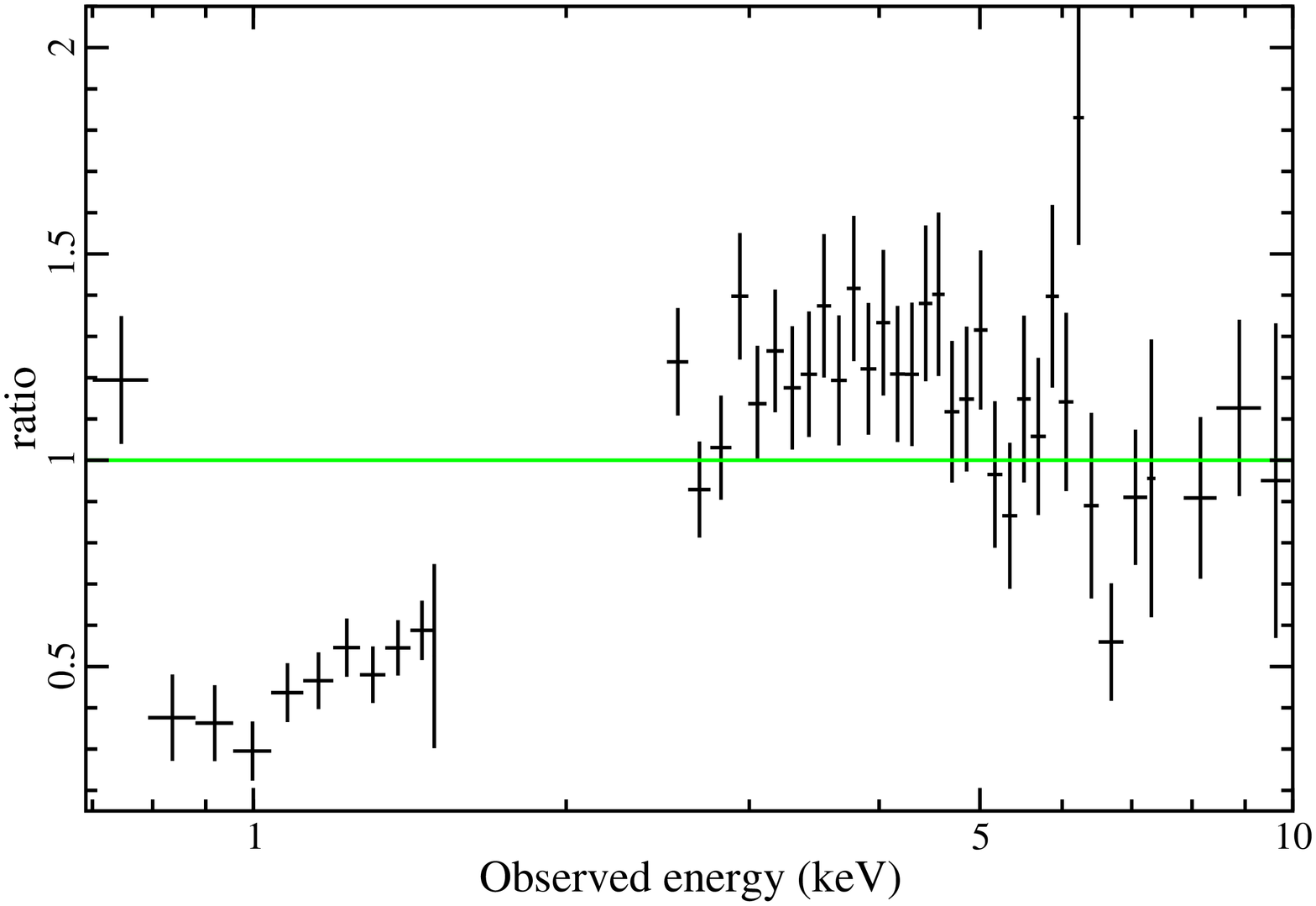}}}
\caption{
Top panel: The fraction variability as a function of the energy is calculated for the entire observation (black open circles) and for all times excluding the flare (red open squares).   The black dashed curve is the expected $\fvar$ if the variability was entirely due to normalisation changes in the power law component.
Middle panel:  The ratio between the two fractional variability spectra (black crosses).  The green line is the constant fit between $2-10\keV$ extrapolated to lower energies.  
Lower panel:  The ratio of the difference spectrum between the flaring periods (when count rate is $>0.6\cps$ in Fig.~\ref{fig:lc}) and non-flaring times ($<0.5\cps$) fitted with a power law ($\Gamma\approx1.6$) between $2-10\keV$.
}
\label{fig:fvar}
\end{figure}

The fractional variability ($\fvar$) in various energy bands is calculated following Edelson \et (2002) with the  \suzaku\ light curves in $5760\s$ bins.  The uncertainties are estimated from Ponti \et (2004).  The $\fvar$ spectrum (Fig.~\ref{fig:fvar}, top panel) exhibits the often seen bell-shaped curve that is interpreted in the blurred reflection model as variable power law and less variable reflection component in a reflection-dominated flux state.  Indeed, the peak amplitude in the $\fvar$ spectrum occurs in the $1.5-3\keV$ band where the power law seems to have the most contribution to the spectrum (Fig.~\ref{fig:Refmean}).
The $\fvar$ spectrum during the 2006 high-flux state was flatter (figure 11 of Larsson \et 2008) than during this epoch, which would be consistent with the spectrum being dominate by a single component.  

The $\fvar$ spectrum also reveals more complex behaviour.  Overplotted on the spectrum is the $\fvar$ expected if only the normalisation of the 
power law was responsible for all the variations.  Clearly, the model indicates that arises from variations in more than one parameter or component.
In addition, in Fig.~\ref{fig:fvar} the fractional variability spectrum is also calculated for times in the light curve that do not include the flare (approximately all of time segment D) (red, open squares).    While this spectrum shows a very similar shape to the one that includes all data (black, open circles), the ratio between the two spectra indicates differences in the soft band below $\sim2\keV$, again reinforcing that changes in power law normalisation alone are not sufficient to describe the spectral variations.

Similar behaviour is apparent in the difference spectrum taken when the source was flaring ($>0.6\cps$ in Fig.~\ref{fig:lc}) and more stable ($<0.5\cps$) (Fig.~\ref{fig:fvar} lower panel).  A power law fitted to the $2-10\keV$ is rather flat ($\Gamma\approx1.6$) and the ratio shows some curvature that resembles the broad reflection component.  When the fit is extrapolated to lower energies there are significant deviations below $1.5\keV$.

\subsection{Principal component analysis in the low-flux state} 
\label{sect:pca}

We now use PCA to investigate variability in the low state data only. The amplitude of the variability is lower, so we have to increase the binning relative to Section~\ref{section_pca_all}. We use 14 broad energy bins to improve the signal to noise, but the analysis is otherwise the same as before.

\begin{figure}
\rotatebox{0}
{\scalebox{0.5}{\includegraphics{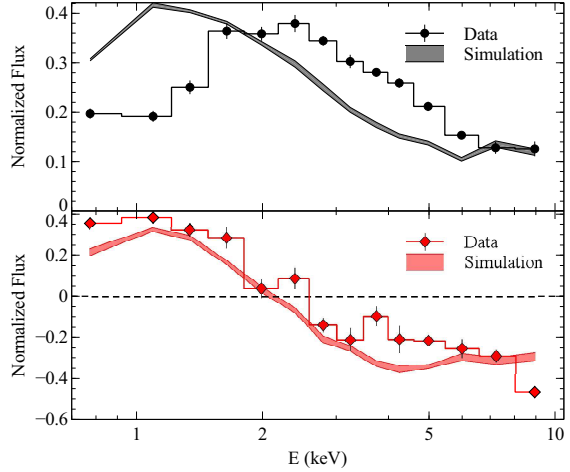}}}
\caption{
First (top, black) and second (bottom, red) principal components returned from PCA of \mrk335. The points show the components returned from the data, and the shaded lines show the results of a simulation of a power law varying in photon index and normalisation, in the presence of a constant reflection component.
}
\label{fig:pca}
\end{figure}

We find two significant components from our analysis of the low state, responsible for $\sim95$ and $\sim1.8$ per cent of the variability respectively, with the remainder indistinguishable from noise. These components are shown in Fig.~\ref{fig:pca}. The primary component is greater than zero at all energies (implying that all bins are correlated), and is suppressed around the energies of the soft excess and iron line. This component is very similar to the RMS variability spectrum shown in Fig.~\ref{fig:fvar}. The second component shows an anticorrelation between low and high energies, indicating the presence of a pivoting effect. These components are qualitatively similar to those found in \mcg\ by Parker \et (2014a), which also showed a primary component suppressed at these energies by a relatively constant reflection component. However, the degree of suppression is much greater in \mrk335, as expected given the higher reflection fraction in this source.

As in Parker \et (2014b), we next compare the principal components from the data with those obtained from simulations. To do this, we simulate a set of 50 30~ks Suzaku spectra based on the best fit reflection model (Table~\ref{tab:Meanfit}) using the \textsc{xspec} command \textsc{fakeit} . We fix all of the parameters apart from the normalisation and photon index of the power law to their best-fit values. We randomly vary the normalisation of the power law by a factor of three for each spectrum, and vary the photon index between 1.7 and 2.2. This set of simulated spectra is then analysed by the PCA code, in exactly the same fashion as the real data. The resulting components are plotted in Fig.~\ref{fig:pca} as shaded regions.  
As with the examination of the $\fvar$ spectrum (Fig.~\ref{fig:fvar}), it is clear that changes in the power law alone are not sufficient to explain the first principal component and suggest that variability in the blurred reflector is likely required.  The model of the second component agrees reasonable well with the data indicating that some fraction of the variability could be attributed to a pivoting power law.

Unlike MCG--6-30-15, we do not find a third order principal component with a correlated iron line and soft excess. This is likely to be a noise issue, as such a component has been found in several other sources which display the same pattern of variability at lower orders (Parker et al., submitted). We note that the level of noise in the Mrk~335 analysis is higher than that in the \emph{XMM-Newton} data of MCG--6-30-15, so we are unable to probe to as high spectral or temporal resolution, and are limited to the first two principal components.

\subsection{Time-resolved spectral variability} 
\label{sec:trsv}

Sections~\ref{sect:rvar} and \ref{sect:pca} indicate that variability from more than one component (i.e. the power law component) is required to describe the rapid spectral variability in the low-flux state of \mrk335.  To examine further, eight spectra are created from the data in each of the $\sim90\ks$ time segments shown in Fig~\ref{fig:lc}.  While this is formally a spectral analysis based on consecutive time slices of the data it also constitutes a flux-resolved analysis given the manner in which the time cuts are done.  Specifically, segments {\it B, F, G, H} represent low-flux levels, segments {\it A, C, E} represent intermediate flux levels, and segment {\it D} corresponds to the high flux level.

The blurred reflection model is applied to each spectrum between $0.7-10\keV$.  The PIN data are not fitted, but the model is extrapolated to higher energies and compared to the average PIN spectrum over the entire observation.  Spectra G and H were found to be very similar and these two consecutive spectra were fitted simultaneously to enhance signal-to-noise. Initially, all the spectra were fitted with a blurred reflection model in which the power law normalisation, photon index, and the ionisation parameter of the reflector were permitted to vary.  All other model parameters were fixed to the average values found in Table~\ref{tab:Meanfit}. 

\begin{figure}
\rotatebox{270}
{\scalebox{0.32}{\includegraphics{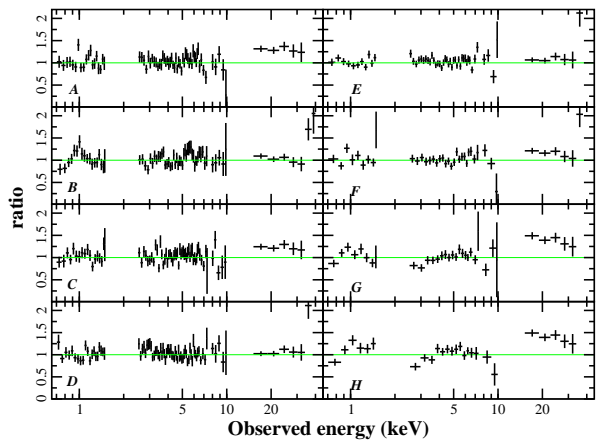}}}
\caption{
The residuals (data/model) shown for time-resolved spectroscopy as defined in Fig.~\ref{fig:lc}.  The blurred reflection model adopted is described in the text.  The variable parameters in each spectral fit are the power law index and normalisation, and the reflector normalisation and ionisation parameter.
The spectra are all reasonable fit with $\redchi\approx 1 - 1.2$ in each case. 
}
\label{fig:timeref}
\end{figure}

The model described depicts a scenario where the reflector responds to changes in the power law illumination with variations in the ionisation parameter.   For half of the spectra this produced an acceptable fit, but in some cases (spectra {\it D, E, G, H}), all during or after the flare, there was significant improvement when the normalisation of the reflector was also allowed to vary.   For these spectra, the addition of one free parameter improved the fit by $\delchi= 27$ for spectra {\it D} and {\it E}, and  $29$ for the combined spectrum {\it G/H}.  The data/model residuals are shown for each spectrum in Fig.~\ref{fig:timeref}.  The spectra are all fitted well ($\redchi\approx 1$) with the described model. Spectra G and H show more significant residuals and enhanced curvature between $2-10\keV$, which could result from the higher background levels in those lowest flux spectra.

\begin{figure}
\rotatebox{270}
{\scalebox{0.32}{\includegraphics{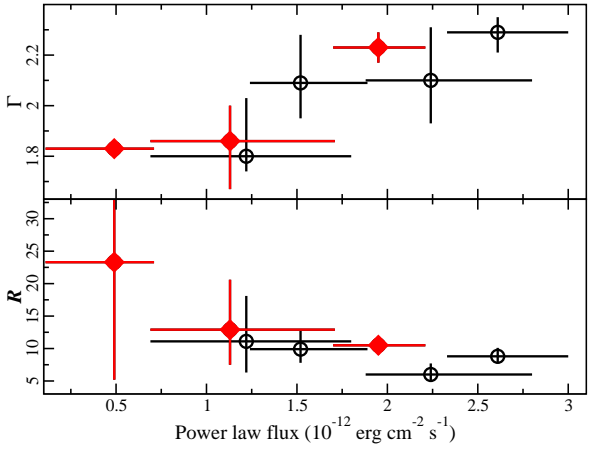}}}
\caption{
Top panel: The power law flux (between $0.1-100\keV$) and photon index are plotted against each other for time-resolved spectral analysis.  The trend between steepening photon index and increasing flux is apparent.  Lower panel:  The reflection fraction ($\mathcal{R}$) plotted against power law flux showing a preference for lower $\mathcal{R}$ at higher continuum flux.  The filled red diamonds identify the post-flare data (segments $E, F, G/H$).
}
\label{fig:pocorr}
\end{figure}

A comparison between some of the parameters derived from the time-resolved spectral analysis reveals relations that are both expected and unexpected if we adopt a light bending interpretation.   The lower panel of Fig.~\ref{fig:pocorr} demonstrates a preference for a lower reflection fraction when the power law flux is higher, which would be anticipated if the bright power law flux is associated with a primary emitter that is more extended over the disc or at a larger distance from the black hole.  In addition, the more general and well-known correlation between the steepening photon index and increasing power law flux (e.g Markowitz \et 2003) is observed in the top panel of Fig.~\ref{fig:pocorr}.

\begin{figure}
\rotatebox{270}
{\scalebox{0.32}{\includegraphics{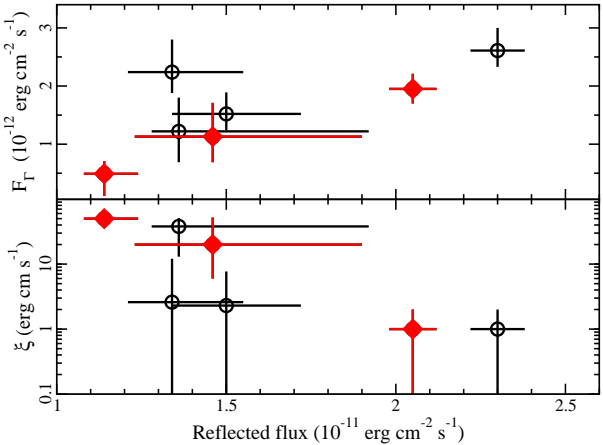}}}
\caption{
The reflected flux (between $0.1-100\keV$) is plotted against the $0.1-100\keV$ power law flux (top panel) and the ionisation parameter ($\xi$).
The filled red diamonds identify the post-flare data (segments $E, F, G/H$).
}
\label{fig:reflcorr}
\end{figure}

In contrast, Fig.~\ref{fig:reflcorr} highlights some of the complexities attributed to \mrk335.  If the AGN is in a low-flux level because it is reflection-dominate then one may expect a clear correlation between the power law and reflected flux.  However, this behaviour is only obviously exhibited in the post-flare data (top panel of Fig.~\ref{fig:reflcorr}).  One would also expect a direct correlation between the reflected flux and ionisation parameter.  In the light bending regime the reflected flux is a better approximation of how much illuminating flux the inner disc is exposed to and this value should be correlated with the ionisation parameter.  However, for \mrk335\ there appears to be an anti-correlation between these parameters suggesting there may be changes in the geometry of the primary source or the nature of the reflector.  We note a need for caution when adopting light bending to interpret these results.  The time bins used in the time-resolved analysis are larger ($\sim90\ks$) than the variability time scales that correspond to the inner few gravitational radii where light bending is most dominant.  However, the complex behaviour is comparable to the behaviour of  \mrk335\ in the intermediate flux state when it was observed with \xmm\ and  more rapid time scales could be probed (G13).

\begin{figure}
\rotatebox{270}
{\scalebox{0.32}{\includegraphics{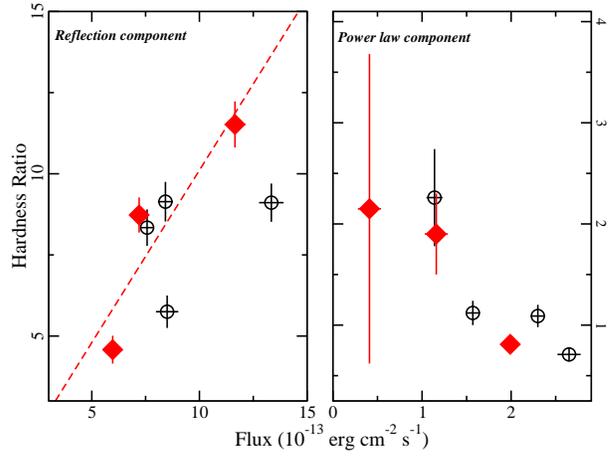}}}
\caption{
The hardness ratio between $7-10\keV$ and $0.7-0.9\keV$ is plotted against the flux in the combined band for the reflection component (left panel) and the power law component (right panel).  
The filled red diamonds identify the post-flare data (segments $E, F, G/H$).
The correlation between $HR$ and flux that is present in Fig.~\ref{fig:hrcr} after the flare seems present in the reflection component, but not the power law component.  The dashed line in the left panel is the best linear fit to the post-flare data (red diamonds).
}
\label{fig:hrcorr}
\end{figure}

The time-resolved spectroscopy does potentially explain the hardness ratio behaviour described in Fig.~\ref{fig:hrcr}.  Using the spectra from each time segment, the hardness ratio between $7-10\keV$ and $0.7-0.9\keV$ is calculated for the reflection and power law component separately, and plotted against the corresponding flux (Fig.~\ref{fig:hrcorr}).  The reflection component appears to exhibit the same behaviour observed in Fig.~\ref{fig:hrcr}. That is, the hardness ratio becomes correlated with the flux only after the flare.  The power law component does not exhibit any obvious trends.  Changes in the ionisation parameter and the normalisation of the reflector are needed to fit the time-resolved spectra after the flare, which could be identifying changes in the disc or corona structure due to the flare.

\section{Discussion} 

\subsection{Long-term variability}
\suzaku\ has observed \mrk335\ on two occasions.  In $2006$ the AGN was in a power-law dominated, bright state.  Seven-years later a triggered observation caught \mrk335\ in a complex, low-flux state.  While the spectra below $10\keV$ were significantly different, the $15-40\keV$ PIN spectra were comparable at both epochs.

Fitting the spectra in a self-consistent manner was difficult with partial covering models that have been previously suggested for this source (e.g. Grupe \et 2008; G12).  The high-flux spectrum could result from a continuum-dominated source without any line-of-sight absorption.  The narrow, neutral iron emission line and some of the $15-40\keV$ flux could be attributed to scattered and reflected emission from the absorbing medium outside the line-of-sight.  However, attempts to reproduce the low-flux spectrum by simply changing the line-of-sight absorption were not successful. 
The addition of two more partial covering absorbers (one ionised and one neutral) improves the fit quality, but still not to the level of the blurred reflection model.  Allowing the primary continuum to vary in addition to the partial covering was also attempted.  While more statistically acceptable fits were obtained,  the models required significant changes in the shape of the power law component (e.g. inverted in some cases)  to account for the significant curvature in the intermediate band (i.e. $\sim 2-6\keV$) and high flux between $\sim 15-40\keV$ that remained constant in the high- and low-flux state.  

Blurred reflection models describe the high- and low-flux spectra rather well and in a self-consistent manner.  The object is clearly power-law dominated during the high state whereas \mrk335\ becomes reflection dominated in the low state when the corona becomes more compact and the source flux diminishes by $\sim10$ in the $0.7-10\keV$ band.  As expected, the power law hardens slightly as its flux drops.
We also considered the possibility the X-ray emission in \mrk335\ arises as described in the blurred reflection model, but that the transition to low state is  attributed to absorption.  Such models were only successful if significant variability of the primary component was also permitted and always resulted in a reflection dominated system.  Such a scenario can not be definitively ruled out given the significant degeneracy arising from over-modelling of the data, but appears rather ad hoc.

Both reflection and partial covering models in the low state require modification by a highly-ionised absorber based on the detection of an absorption-like feature at about  $7\keV$.   The feature could be attributed to the hot medium in the multi-zone warm absorber detected in \mrk335\ during the 2011 observation when the source was in an intermediate-flux state (L13).  In fact, including only the hottest component of the multi-zone warm absorber found in 2011 with the same parameters used by G13 significantly improved the reflection model.   The partial covering model exhibited a much stronger feature requiring the column density of the hot medium to have become $10\times$ higher than it was in $2009$.

The blurred reflection scenario appears to be a simpler model, but still reveals some challenges.  When the high- and low-flux spectra are fitted separately both prefer a different combination of iron abundance and black hole spin parameter even though all other parameters are comparable between the two flux states.  The fit degeneracy between iron abundance and spin has been recognised in a number of previous works (e.g. Reynolds \et 2012; Walton \et 2013).  The challenge in modelling two very different looking spectra with the same parameters is brought to the forefront in this work.  However, it is important to note that for both flux states the measurements favour an iron overabundance and a nearly maximum black hole spin.  Simulations by Dauser \et (2014) seem to support the higher spin value measured during the low-flux state ($a=0.995$).  Dauser \et argue that high reflection fractions in excess of $2$ can only be achieved for rapidly spinning black holes.  The reflection fraction of $\mathcal{R}\approx 8$ in the low-flux observation requires a spin parameter $a>0.96$  assuming the accretion disc extends down to the ISCO, which is inconsistent with the lower value of $\sim0.93$ measured in the high-flux state.   Conservatively, if we were to consider the combined range of possible values measured from both independent fits \mrk335\ requires an iron abundance that is a factor of $\sim 2-7$ above solar and a spin parameter $a> 0.91$.  The results are consistent with previous measurements of \mrk335\ using data from different telescopes and with the AGN in different flux states (e.g. Parker \et 2014; G13; Walton \et 2013).

The reflection fraction during the high-flux observation is extremely low and inconsistent if considering a compact, isotropic emitting source irradiating an infinite plane-parallel disc.  One could consider the possibility that the primary source is perhaps the base of a jet that is moving away from the disc at mildly relativistic speeds (e.g. Reynolds \& Fabian 1997; Beloborodov 1999).  With the primarily emission being beamed away from the disc the observer would see a lower reflection fraction and brighter source.  The situation has been proposed to describe the low reflection fraction in Cyg~X-1 (Beloborodov 1999) as well as other radio-quiet Seyfert galaxies like \mrk335\ (e.g. Gallo \et 2011b).  From equation (3) of Beloborodov (1999), 
$\mathcal{R} = (1- \beta/2)(1-\beta \cos i)^3 / (1+\beta)^2$,
the bulk velocity of the illuminating blob of material ($\beta=v/c$) is related to $\mathcal{R}$, where $i$ is the inclination
of the system.  Adopting $i=60\deg$ and $\mathcal{R}=0.34$ as found in Sect.~\ref{sect:meanref} renders $\beta \approx 0.28$.  
The velocity is similar to what was reported by Liu \et (2014) who suggest the X-ray emission in some AGN can be attributed to a bipolar outflowing corona with bulk velocities of $\beta \approx 0.3-0.5$.  Since \mrk335\ is radio-quiet the material likely does not escape (e.g. Ghisellini \et 2004), which would imply the launch point is within approximately $26\rg$ from the black hole.   The result is comparable to the height of the source above the accretion disc that was measured in Section~\ref{sect:meanref} ($h=23^{+6}_{-4} \rg$).

The jet interpretation during the high-flux levels is consistent with the analysis of the emissivity profile.  In the high-flux observation the emissivity profile falls off steeply to about $3\rg$ at which point the profile flattens considerable to $r^{-2}$ over the extent of the disc.  Wilkins \& Fabian (2012) note that such a profile is consistent with coronal emission that is being collimated away from the disc.  From the spectral analysis, the putative jet should be confined to within $25\rg$ if the material does not escape the system.  Longinotti \et (2007) report possible inflowing material in an \xmm\ observation of \mrk335\ that could be consistent with matter falling back down on the disc.  \mrk335\ is radio-quiet, but not radio-silent thus it is possible that material could be ejected from the system on occasion.   We note the emissivity profile shown in Fig.~\ref{fig:Emiss} for the high state indicates there is significant illumination of the inner accretion disc.  This emphasizes the difficulty in describing the primary X-ray source as a point source, and implies that it likely has some vertical extent. 

The spectral modelling in Sect.~\ref{sect:meanref}, shows that the contribution of the power law component to the spectrum has diminished by a factor of ten in the low flux observation, and the spectrum is now reflection-dominated. 
A possible explanation of the lags in Fig.~\ref{fig:lag} is that the continuum lags have been mirrored by the reflection component. In this picture, hard lags in the power-law continuum can be thought of as a series of soft flares followed, after some delay, by hard flares. The soft flares produce a relatively soft reflection spectrum while the hard flares produce a relatively harder reflection spectrum (see for example figure 6 in Garc\'ia \et 2013). The result is the reflection spectrum itself as a whole will have hard lags.

\subsection{\mrk335\ in the low-flux state}

The 2013 \suzaku\ data mark the longest observation of \mrk335\ in a low-flux state and permit the most detailed analysis of its behaviour at low-flux levels.  
Various lines of study used in this work (hardness ratios, $\fvar$, principal component analysis) support the reflection dominated interpretation for the NLS1.  Analysis of the emissivity profile indicate the corona is compact during the low-flux extending out only to about $5\rg$.  The variability in the low state is mostly driven by the power law component, but not exclusively, and variability of the blurred reflection emission is also required. 

The flare-like feature in the middle of the light curve seems to mark a transition in the variability behaviour of \mrk335.  Specifically, one hardness ratio ($7-10\keV / 0.7-0.9\keV$) becomes correlated with brightness only after the flare.   Interestingly, both energy bands should be dominated by the reflection component.  Slicing the data into consecutive time-bins of approximately $90\ks$ duration, the time-resolved spectra were examined to study the time evolution of various parameters.  One particular relation that stands out after the flare, but not obviously prior to it, is an anti-correlation between the disc ionisation parameter and the reflected flux (or alternatively the reflection fraction).  Essentially, as the flux increases the soft excess diminishes because the ionisation parameter increases.  Taken at face value, it was shown in Fig.~\ref{fig:reflcorr} that this behaviour could account for observed post-flare correlation between hardness ratio and count rate.
The change in the shape of the soft-excess could also explain the discrepancies in the PCA (Fig.~\ref{fig:pca}) and $\fvar$ (Fig.~\ref{fig:fvar}) model fits.  Similar behaviour was observed in another NLS1, I~Zw~1 (Gallo \et 2007) indicating that such behaviour may not be rare.

The origin of the behaviour is not obvious.  The reflected flux should indicate the level of power law illumination seen by the disc.  As more light shines on the disc, more light should be reflected off the disc, but also the ionisation parameter should increase.  Thus, one would expect the reflected flux and ionisation parameter to be correlated.  The anti-correlation possibly triggered by the flare could suggests something about the disc structure.  If the flare is associated with some ejection of disc material then this could temporarily change (i.e. decrease) the density of the inner disc and raise the ionisation parameter.
Alternatively, and more simply,  if the primary emitter becomes more compact the ionising flux would concentrate on the inner part of the disc making that region highly ionised, but the average ionisation parameter that is measured would be low.

\section{Conclusions } 

We have reported the deepest X-ray observation of the NLS1 \mrk335\  in the low-flux state and made comparison with the source when it was $\sim 10\times$ brighter in 2006.  Describing the high- and low-flux spectra self-consistently seems challenging with partial covering models.  Blurred reflection from an accretion disc around a nearly maximum spinning black hole appears more likely and is consistent with long-term and rapid variability.

The high-flux \suzaku\ observation of \mrk335\ is consistent with continuum-dominated, jet-like emission.   It can be argued that the ejecta must be confined to within $\sim25\rg$ if it does not escape the system.   Longinotti \et (2007) have made claims of high-velocity inflowing material in an \xmm\ observation of \mrk335, which could be consistent with the material not reaching the escape velocity and raining back down toward the disc.   Simultaneous X-ray and radio monitoring of \mrk335\ to investigate jet activity would be fruitful.

During the low-flux the corona becomes compact and only extends to about $5\rg$ from the black hole, and the spectrum becomes reflection-dominated.  The variability is dominated by the power law component, but changes in the reflector (primarily the ionisation parameter) over time are required.


\section*{Acknowledgments}

Many thanks to the \suzaku\ team for accommodating the target-of-opportunity.
We thank the referee for constructive comments that helped improve the paper.
LCG would like to thank C. S. Reynolds for helpful discussion and A. Golob for assistance with preparing figures.
DRW is supported by a CITA National Fellowship.




\bsp
\label{lastpage}
\end{document}